\documentclass{article}

\usepackage{PRIMEarxiv}

\usepackage[utf8]{inputenc} % allow utf-8 input
\usepackage[T1]{fontenc}    % use 8-bit T1 fonts
\usepackage{hyperref}
\usepackage{url}            % simple URL typesetting
\usepackage{subcaption} % for subfigure environment
\usepackage{float}      % for [H]
\usepackage{booktabs}       % professional-quality tables
\usepackage{bm}
\usepackage{amsfonts}       % blackboard math symbols
\usepackage{amsmath}
\usepackage{amssymb}
\usepackage{nicefrac}       % compact symbols for 1/2, etc.
\usepackage{microtype}
\usepackage{cleveref}
\usepackage{xcolor}
\usepackage{mathtools}
\usepackage{algpseudocode}
\usepackage{algorithm}
\usepackage{color,soul}
\usepackage{caption}
\usepackage{lipsum}
\usepackage{fancyhdr}       % header
\usepackage{graphicx}       % graphics

\graphicspath{{media/}}     % organize your images and other figures under media/ folder
\usepackage{enumitem}
\makeatletter
\newcommand{\printfnsymbol}[1]{%
  \textsuperscript{\@fnsymbol{#1}}%
}

\makeatother

\title{TOMATOES: Topology and Material Optimization for Latent Heat Thermal Energy Storage Devices}

\author{ \and
Rahul Kumar Padhy \\
Department of Mechanical Engineering\\
University of Wisconsin-Madison\\
Madison, WI, USA \\
\texttt{rkpadhy@wisc.edu}
\and 
 Krishnan Suresh \\
 Department of Mechanical Engineering\\
 University of Wisconsin-Madison\\
 Madison, WI, USA \\
\texttt{ksuresh@wisc.edu} \\
\and 
 Aaditya Chandrasekhar \\
 Department of Mechanical Engineering\\
 Northwestern University\\
 Evanston, IL, USA \\
\texttt{aadityacs@northwestern.edu} \\
}

\begin{document}
\maketitle

\begin{abstract}

Latent heat thermal energy storage (LHTES) systems are compelling candidates for energy storage, primarily owing to their high storage density. Improving their performance is crucial for developing the next-generation efficient and cost effective devices. Topology optimization (TO) has emerged as a powerful computational tool to design LHTES systems by optimally distributing a high-conductivity material (HCM) and a phase change material (PCM). However, conventional TO typically limits to optimizing the geometry for a fixed, pre-selected materials. This approach does not leverage the large and expanding databases of novel materials. Consequently, the co-design of material and geometry for LHTES remains a challenge and unexplored.

To address this limitation, we present an automated design framework for the concurrent optimization of material choice and topology. A key challenge is the discrete nature of material selection, which is incompatible with the gradient-based methods used for TO. We overcome this by using a data-driven variational autoencoder (VAE) to project discrete material databases for both the HCM and PCM onto continuous and differentiable latent spaces. These continuous material representations are integrated into an end-to-end differentiable, transient nonlinear finite-element solver that accounts for phase change. We demonstrate this framework on a problem aimed at maximizing the discharged energy within a specified time, subject to cost constraints. The effectiveness of the proposed method is validated through several illustrative examples.

  \begin{figure}
 	\begin{center}
		\includegraphics[scale=0.4,trim={0 0 0 0},clip]{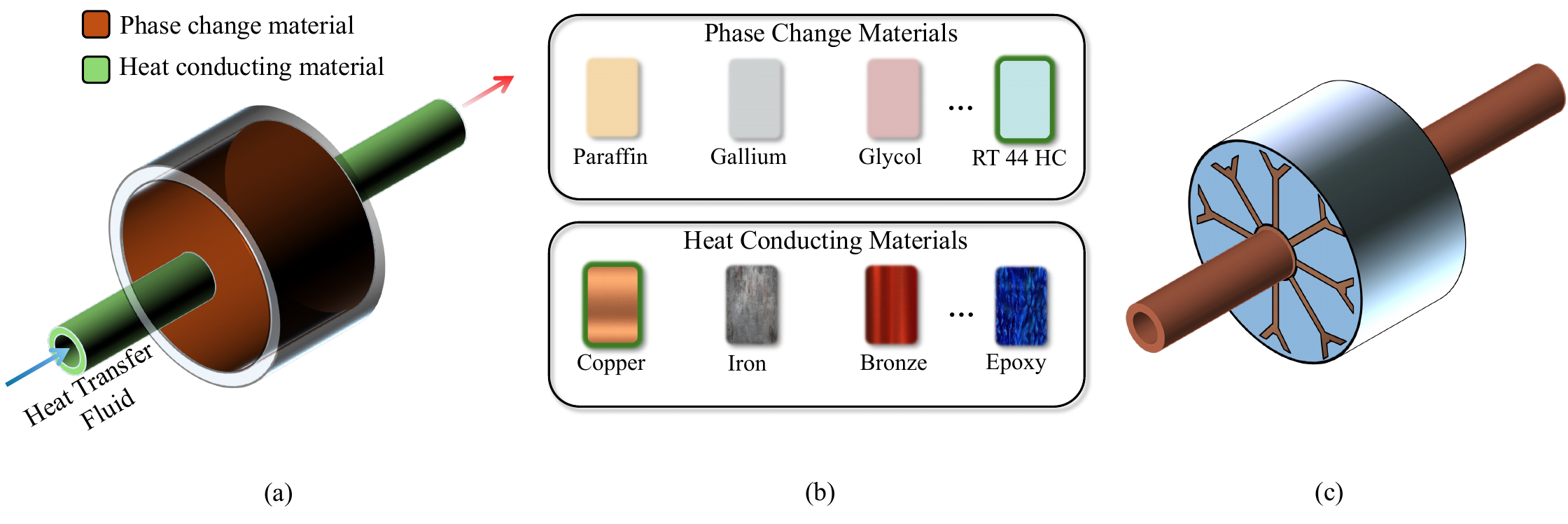}
 		\caption{Graphical abstract of the co-design framework. Given (a) a latent heat thermal energy storage (LHTES) device and, (b) databases of candidate phase change materials (PCMs) and heat conductivity materials (HCMs) the framework co-optimizes (c) the material selection and distribution of the PCM and HCM.}
        \label{fig:problem}
	\end{center}
 \end{figure}

\end{abstract}


\section{Introduction}
\label{sec:intro}
 
The imperative for decarbonization \cite{carnie2024decarbonising}, grid stabilization \cite{aftab2021phase}, and waste-heat recovery \cite{huang2023performance} has increased the demand for advanced energy storage technologies. Among competing technologies, latent-heat thermal energy storage (LHTES)—which stores energy in the latent heat of phase-change materials (PCMs)—has emerged as a promising candidate due to its high storage density \cite{sharma2005latent}. The realization of next-generation LHTES systems is seen as critical, supported by advances in both materials and system design.

On the materials front, advances in the discovery and design of materials have expanded the palette of energy-storage and transport materials, with improved storage capacity, thermal conductivity, and stability \cite{khare2012selection, sharma2005latent}. In parallel, new manufacturing methods—particularly additive and hybrid manufacturing—have enabled the fabrication of complex heat-conduit geometries that can substantially enhance thermal transport within LHTES systems \cite{freeman2023advanced}. Together, these developments create both opportunity and complexity for the practicing engineer in designing optimal LHTES systems.

In particular, effectively leveraging this expanded material and geometric design space to extract optimal performing configurations presents a central challenge \cite{chandrasekhar2022integrating}. Designers must select from a vast and diverse materials space while accounting for the system's nonlinear, transient, and strongly coupled thermophysical behavior. Furthermore, the geometric freedom afforded by modern manufacturing surpasses the capabilities of traditional, intuition-driven design workflows. Consequently, these manual design approaches are limited in scope and frequently overlook high-performance configurations.

To overcome these limitations, computational methods offer a systematic path forward. Specifically, topology optimization (TO) \cite{sigmund2013_TO_review, duysinx1998topology} and machine learning (ML) \cite{woldseth2022TO_ML_review} have emerged as powerful tools for automating design exploration. TO employs gradient-based optimization to navigate complex design landscapes, systematically discovering novel geometries and material layouts to improve performance while respecting governing physics and design constraints. Complementing this physics-driven approach, ML methods excel at discovering, organizing, and navigating expansive material databases to inform design selection \cite{chandrasekhar2022integrating}.

The integration of TO and ML is well-positioned to improve the performance of LHTES devices. A critical step toward this goal is to address the strong coupling between geometry and material properties, which necessitates their simultaneous optimization. Hitherto, applications of TO have largely ignored this coupling, focusing primarily on geometric optimization while fixing the material choice \textit{a priori} \cite{pizzolato2017LHTES, pizzolato2017TO_PCM, wang2023topology}. 
For instance, consider a vertical tube carrying a cold heat-transfer fluid (HTF), which is surrounded by both a low-conductivity PCM and a high-conductivity material (HCM) as shown in \cref{fig:problem}(a). In traditional approaches, the PCM and HCM are first selected a priori from a material database (\cref{fig:problem} (b)), and their spatial distribution (\cref{fig:problem}(c)) is subsequently optimized using the prescribed material properties. Consequently, the more challenging problem of co-designing geometry and material has received far less attention.

A primary barrier to this simultaneous optimization is that material selection is inherently discrete, whereas gradient-based topology optimization requires a differentiable design space. This fundamental incompatibility between discrete material choice and continuous geometric optimization makes true co-design—the joint optimization of material and geometry—exceptionally challenging.

This paper directly addresses this methodological gap. Building upon recent work \cite{chandrasekhar2022integrating}, we introduce an ML-enabled framework that bridges the discrete and continuous domains. Our approach employs a variational autoencoder (VAE) \cite{kingma2013auto} to embed discrete material candidates into a continuous and differentiable latent space. This learned material representation is then integrated directly into a density-based topology optimization formulation, enabling the simultaneous co-design of material and geometry within a single, unified framework. We demonstrate that this co-design methodology yields LHTES devices with performance superior to both conventional baseline designs and sequential optimization workflows.

\section{Related Work}
\label{sec:relatedWork}

This section surveys the state of the art in the design of LHTES devices. The survey is structured around the central contribution of this paper: the concurrent material selection and topology optimization through an ML augmented framework for the design of LHTES devices. We organize the review into three subsections. The first subsection reviews TES devices and the materials commonly employed therein. The second subsection surveys the use of TO in thermal management, in particular LHTES devices. The third subsection examines the co-design of material and geometry, outlining the limitations of conventional approaches and recent developments that incorporate machine learning.

\subsection{Thermal Energy Storage}
\label{sec:relatedWorkTES}

The rapid growth of renewable energy and the increasing variability of supply and demand highlight the need for next generation energy storage systems \cite{sarbu2018comprehensive}. A wide range of storage technologies has been developed to address these challenges, including electrochemical systems \cite{yang2011electrochemical}, mechanical storage \cite{alami2020mechanical}, and TES \cite{sarbu2018comprehensive}, each with distinct operating principles, efficiency levels, costs, and environmental impacts \cite{gur2018review, rahman2022recent}. Among these, TES plays a vital role in applications involving heat management and the integration of renewable energy \cite{rehman2015pumped}. TES systems are typically classified into three categories: sensible heat, latent heat, and thermochemical storage \cite{sarbu2018comprehensive}. Sensible storage, while widely deployed in hot water tanks and packed-bed systems, generally exhibits lower storage densities compared to latent and thermochemical TES types \cite{sarbu2018comprehensive, zhang2016thermal}. Furthermore, thermochemical storage can achieve high energy density, but it relies on reversible chemical reactions that require high material stability and introduce significant complexity in terms of kinetics and long-term durability \cite{bauer2021fundamentals, gur2018review}. Latent heat storage combines high storage density with nearly isothermal charging and discharging, making it well suited for renewable integration and thermal management applications \cite{sarbu2018comprehensive, pizzolato2017TO_PCM, tian2021bionic}.

LHTES systems rely on PCMs as the medium for storing and releasing energy \cite{pizzolato2017LHTES}. The performance of an LHTES device, therefore, depends critically on the type of PCM employed and how it is integrated with HCMs for heat transfer. Broadly, PCMs are categorized into three groups: organics, inorganics, and eutectics \cite{sarbu2018comprehensive}. To begin with, organic PCMs , such as paraffins and fatty acids, are widely used because they are chemically stable, non-corrosive, and readily available \cite{sarbu2018comprehensive}. However, their low thermal conductivity often limits heat transfer rates, which necessitates the inclusion of HCMs such as copper or aluminum to improve performance \cite{Ashby1993MaterialSelection, mehmood2018materialSelectionMEMSAshby, pizzolato2017TO_PCM}. In contrast, inorganic PCMs, particularly salt hydrates, offer higher volumetric storage capacity and thermal conductivity. Yet, they are frequently affected by drawbacks such as supercooling, phase segregation, and material compatibility issues that hinder long-term reliability \cite{sarbu2018comprehensive}. Alternatively, eutectic PCMs combine two or more components in fixed ratios, producing well-defined melting points tailored to specific operating conditions \cite{sarbu2018comprehensive}. This makes them suitable for applications that require precise temperature control. This diversity of PCM options and associated trade-offs motivates the need for advanced design approaches; the next subsection reviews the use of TO as a methodology for designing efficient LHTES systems.

\subsection{Topology Optimization of Thermal Energy Storage Devices}
\label{sec:relatedWorkTO}

TO has transformed the design of structural \cite{sigmund2013_TO_review}, fluid \cite{borrvall2003topology, padhy2024fluto, padhy2025toflux}, photonics \cite{christiansen2021inverse, padhy2024photos}, and thermal systems \cite{tian2004effects}, enabling automated generation of high-performance geometries that outperform traditional, manually designed configurations. Its relevance has increased further with the advent of additive manufacturing, which enables the fabrication of the complex, non-intuitive geometries produced through optimization studies \cite{ge2020additive}. In thermal management applications, TO has been employed to enhance the performance of heat sinks, heat exchangers, and cooling plates, producing designs that consistently outperform conventional fin-and-tube solutions in terms of thermal resistance, compactness, and efficiency \cite{mo2021topology, careri2023additive}.

LHTES design using TO was introduced by the seminal work of \cite{ pizzolato2017LHTES}, who showed that optimized fin arrangements within PCM domains could yield branching structures that significantly improved melting and solidification rates. Subsequent studies extended this framework to multi-tube configurations \cite{pizzolato2017TO_PCM}, bio-inspired fin geometries \cite{tian2021bionic}, and PCM-based heat sinks whose topology-optimized designs were experimentally validated \cite{see2022experimental}. More recent contributions have examined battery thermal management systems incorporating PCMs \cite{mo2021topologyPCMCoolingPlates}, and transient PCM-integrated heat sinks \cite{iradukunda2020transient}. Furthermore, \cite{xiaoping2023LHTES} applied TO to HCM/PCM composites with variable HCM volume fraction under natural convection to optimize charging behavior. Despite these advances, most TO studies select material properties for PCM and HCM a priori. \cite{pizzolato2020LHTES} showed that this sequential choice of materials and geometry is suboptimal, and demonstrated that material selection can strongly influence the optimal fin topology. However, their approach was restricted to a small, discrete set of materials, thereby limiting generality and preventing exploration of the broader material design space. Addressing this limitation requires frameworks that can represent material parameters in a continuous manner, enabling geometry and material to be optimized together.

\subsection{Material–Geometry Co-Design}
\label{sec:relatedWorkMTO}
The challenge of simultaneous material–geometry optimization has a long history in mechanical design. Ashby’s charts and indices provided intuitive performance metrics for simple cases \cite{ashby2011MaterialSelection, Ashby1993MaterialSelection} and remain influential in early-stage design.  Yet, these methods are inadequate for multiphysics problems such as LHTES, where geometry and material choices are strongly coupled, and the governing physics is nonlinear and transient \cite{pizzolato2020LHTES}.

To address these challenges, researchers explored smooth material functions \cite{ananthasuresh2003coGeomMaterial}, hybrid approaches such as genetic algorithms and sequential quadratic programming (SQP) \cite{roy2021hybrid} and mixed-discrete formulations \cite{achleitner2022materialSelectionCompliant}. While these methods provide valuable insights, they often incur high computational costs and limited scalability \cite{chandrasekhar2022integrating, nocedal1999Optimization}. In LHTES, similar limitations have restricted co-design to the evaluation of a handful of candidate materials \cite{pizzolato2020LHTES}.

Despite these advances, reliance on discrete material sets and the high computational cost continue to limit progress in material–geometry co-design. ML has recently emerged as a way to address these limitations. \cite{woldseth2022TO_ML_review, chandrasekhar2021tounn} demonstrated that neural networks can serve as surrogate models \cite{padhy2025voroto} or design representations in topology optimization, enabling efficient exploration of high-dimensional design spaces. Building on this direction, \cite{chandrasekhar2022integrating} introduced the use of VAEs to embed discrete material databases into continuous and differentiable latent spaces. A VAE is a type of generative neural network that learns a compressed latent representation of input data while ensuring continuity and differentiability, allowing discrete material properties to be represented in forms amenable with gradient-based optimization \cite{kingma2013auto}. The present work extends the VAE-based co-design framework to LHTES, where phase change exhibits nonlinear and transient behavior. By integrating a VAE-trained decoder with a finite element solver for phase-change heat transfer, our framework enables the simultaneous optimization of material and geometry within an end-to-end differentiable environment.

The remainder of this paper is structured as follows. The governing physics and geometry representation are presented in \Cref{sec:method_govEq} and \Cref{sec:method_designRepMatInterp}. \Cref{sec:method_material} describes the VAE framework for material embedding. In \Cref{sec:method_optimization}, we formulate the co-design optimization problem where the VAE is integrated with geometry optimization, leveraging automatic differentiation to compute sensitivities. \Cref{sec:expts} demonstrates the effectiveness of this framework on several LHTES benchmark problems, showing that simultaneous optimization of material and geometry yields superior performance compared to optimizing either in isolation. Finally, \Cref{sec:discussion} summarizes the contributions and outlines future directions.

\section{Proposed Method}
\label{sec:method}

\begin{figure}[H]
 	\begin{center}
		\includegraphics[scale=0.45,trim={0 0 0 0},clip]{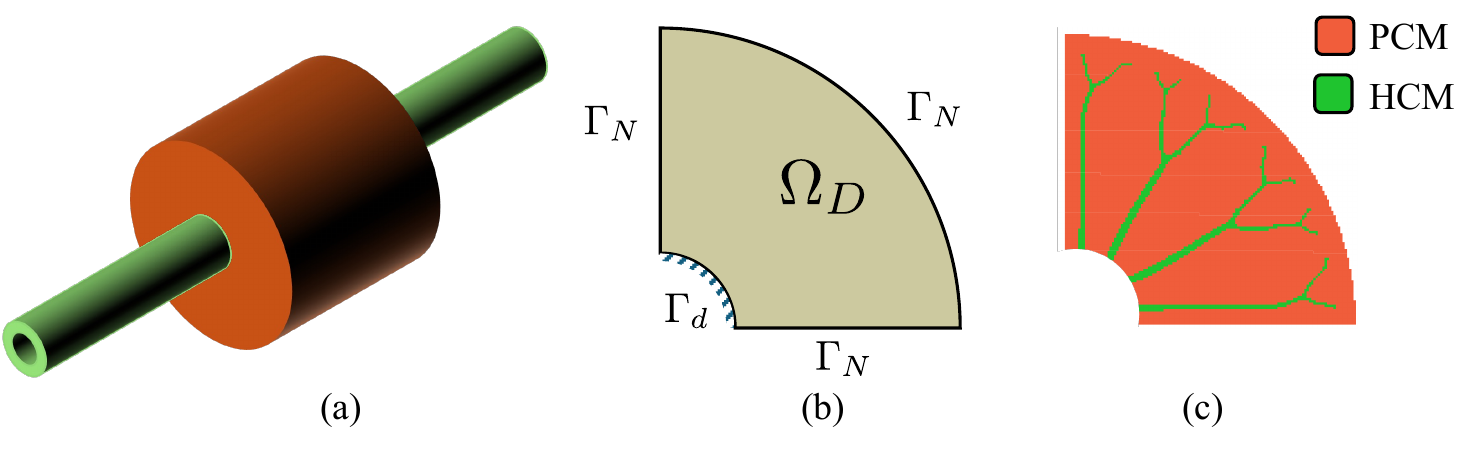}
        \caption{The (a) shell-tube LHTES configuration, (b) the corresponding axisymmetric design domain, and (c) a representative optimized material distribution. }
        \label{fig:domain_bc}
	\end{center}
 \end{figure}

This study addresses the co-optimization of topology and material selection for latent heat thermal energy storage (LHTES) devices. We begin the formulation by assuming a prescribed design domain and boundary conditions, such as the axisymmetric annular configuration shown in \Cref{fig:domain_bc}, which is typical of thermal storage systems. Furthermore, we assume the availability of a material database containing candidate high-conductivity materials (HCMs) and phase change materials (PCMs). This database includes relevant properties such as thermal conductivity, specific heat capacity, and mass density for all materials, cost per unit mass for HCMs, and latent heat of fusion and melting temperature for PCMs.

Conventionally, the two material classes have distinct and complementary thermal properties. HCMs, typically metals, exhibit high thermal conductivity, whereas PCMs, often paraffins, offer superior thermal storage capacity through both sensible and latent heat but have low thermal conductivity. The HCMs function as thermal conduits to distribute heat into the PCM. This establishes a natural trade-off between the system's total heat-storage capacity and its ability to transfer heat effectively into and out of the system.

The design objective is then to determine the optimal spatial distribution and selection of HCM and PCM. In particular, this work extends prior geometric optimization studies by concurrently optimizing the material selection for both constituents from the database. We employ gradient-based optimization to maximize the energy discharged over a given period, subject to a cost constraint on HCM use.

The governing PDE and its finite element formulation are discussed in \Cref{sec:method_govEq,sec:method_fea} respectively. The material distribution in the design is represented using conventional density-based methods employed in topology optimization (TO), as detailed in \Cref{sec:method_designRepMatInterp}. To make the discrete material database amenable to gradient-based optimization, we first convert it into a continuous, differentiable form using a Variational Autoencoder (VAE), as detailed in \Cref{sec:method_material}. The decoder component of this model is then employed within the gradient-based optimization algorithm, as discussed in \Cref{sec:method_optimization}. A complete summary of the optimization algorithm is provided in \Cref{sec:method_algorithm}.

\subsection{Governing Equations}
\label{sec:method_govEq}

This study focuses on the topology optimization of the thermal energy discharge process in a two-dimensional domain composed of an HCM and a PCM. The computational domain, $\Omega_D$, represents a quarter cross-section of the thermal energy storage (TES) unit (\Cref{fig:domain_bc}(b)). This domain eventually is partitioned, through TO, into two non-overlapping subdomains, $\Omega_{HCM}$ and $\Omega_{PCM}$, such that their union forms the complete design domain: $\Omega_D = \Omega_{HCM} \cup \Omega_{PCM}$.

The domain is initialized with a uniform temperature $T_I$. For the discharge process, a constant temperature $T_d$ is prescribed on the internal boundary $\Gamma_d$, simulating contact with a cold heat-transfer fluid (HTF). The solidification process within the PCM is assumed to be dominated by heat conduction, with convection neglected. \textcolor{black}{While natural convection can enhance heat transfer rates in the liquid phase, particularly during long durations \cite{wang2023topology}, we adopt this conduction-dominated model to establish the foundational framework for simultaneous material-geometry co-design of LHTES systems.} Thus, to model the heat transfer behavior, the transient heat conduction equation is employed. The governing equation, along with the associated boundary and initial conditions, can be expressed as:

\begin{equation}
\label{eq:residual_form}
R(T) \coloneqq
\begin{cases}
  \rho \frac{\partial(\breve{c}(T) T)}{\partial t} - \frac{\partial}{\partial x_i}\left(k_{ij}\frac{\partial T}{\partial x_j}\right) = 0 & \text{for } \mathbf{x} \in \Omega_D, \: \forall t \\
  \\
  T - T_d = 0 & \text{for } \mathbf{x} \in \Gamma_d, \: \forall t \\
  \\
  \left(-k_{ij}\frac{\partial T}{\partial x_j}\right) \cdot \mathbf{n} = 0 & \text{for } \mathbf{x} \in \Gamma_N, \: \forall t \\
  \\
  T(\mathbf{x}, t) - T_I(\mathbf{x}) = 0 & \text{for } \mathbf{x} \in \Omega_D, \: t = 0
\end{cases}
\end{equation}

\begin{figure}[H]
 	\begin{center}
		\includegraphics[scale=0.27,trim={0 0 0 0},clip]{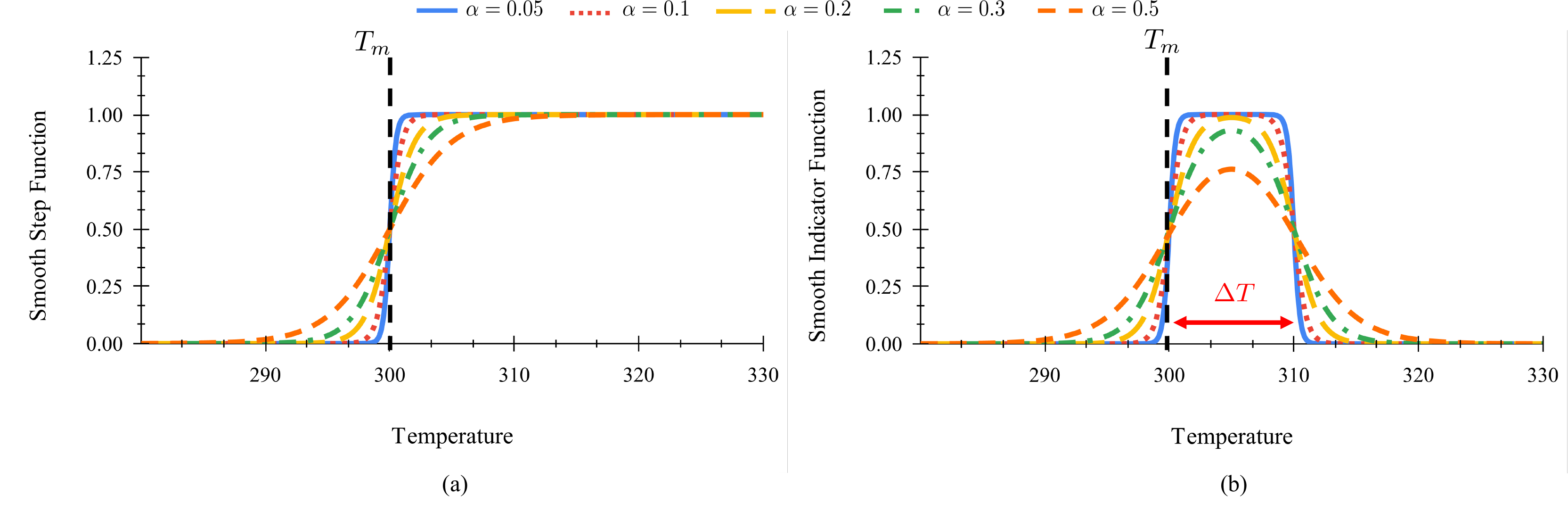}
        \caption{(a) Smooth step function and (b) indicator function.}
        \label{fig:smooth_step}
	\end{center}
 \end{figure}

Here, $T$ is the temperature field, $\rho$ is the density, $\breve{c}$ is the effective specific heat capacity, $k_{ij}$ is the isotropic thermal conductivity tensor. The Neumann boundary condition on $\Gamma_N$ is considered to be zero to enforce the adiabatic and symmetry conditions, where $\mathbf{n}$ is the outward-pointing normal vector. To account for phase change, we adopt the apparent heat capacity method \cite{pizzolato2017LHTES}, where the material's heat capacity $c_p$, is augmented with the latent heat of fusion, $L$, to arrive at the temperature-dependent effective specific heat capacity $\breve{c}(T)$:

\begin{equation}
\breve{c}(T) = c_p + \frac{L}{\Delta T} \cdot \Pi(T)
\end{equation}

where the smooth indicator function, $\Pi(T)$, approximates a rectangular pulse (\Cref{fig:smooth_step}(b)) active within the mushy zone temperature range $[T_m, T_m + \Delta T]$ and is constructed as the difference between two smooth step functions \cite{pizzolato2017LHTES}:

\begin{equation}
\Pi(T) = \Psi(T; T_m, \alpha, \Delta T) - \Psi(T; T_m + \Delta T, \alpha, \Delta T)
\end{equation}

where, each smooth step function, $\Psi(T; T_c, \alpha, \Delta T)$ as illustrated in \Cref{fig:smooth_step}(a), provides a differentiable approximation of the Heaviside step function:

\begin{equation}
\Psi(T; T_c, \alpha, \Delta T) = \frac{1}{2} \left[ 1 + \tanh\left(\frac{T - T_c}{\alpha \Delta T}\right) \right]
\end{equation}
 
where, $T_c$ represents the center of the transition; at the melting solidus temperature $T_m$ and the melting liquidus temperature $T_m + \Delta T$. The parameter $\alpha \;(=0.25 \text{ in our experiments})$ is a dimensionless parameter that controls the transition width as a fraction of the mushy zone width, $\Delta T$. The PDE in \Cref{eq:residual_form} is spatially discretized and solved using standard finite element procedures as detailed in \Cref{sec:method_fea}. The subsequent section details the parameterization of the constituent material for each element.

%------------------------------------%

\subsection{Design Representation and Material Interpolation}
\label{sec:method_designRepMatInterp}

We employ standard element-wise pseudo-densities \cite{sigmund200199} to represent the design. In particular, the pseudo-density of element $e$ is $\gamma_e \in [0,1]$, where a value of zero indicates the element is composed of the HCM, and a value of one indicates the element is composed of the PCM. Observe that although the pseudo-densities may assume any continuous value in [0,1], we require a binary design at convergence. To this end, we use a material interpolation scheme that penalizes intermediate densities. In particular, the effective thermal conductivity is obtained using the standard Solid Isotropic Material with Penalization (SIMP) scheme \cite{pizzolato2017LHTES}:
\begin{equation}
k_{eff}(\gamma) = k_{HCM} + (k_{PCM} - k_{HCM})\gamma^p
\label{eq:interpolation_conductivity}
\end{equation}
where we set $p$ using a continuation scheme in our experiments. The specific heat capacity $(c)$, mass density $(\rho)$, and latent heat $(L)$ are interpolated linearly:

\begin{equation}
c_{eff}(\gamma) = c_{HCM} + (c_{PCM} - c_{HCM})\gamma
\label{eq:interpolation_specificHeat}
\end{equation}

\begin{equation}
\rho_{eff}(\gamma) = \rho_{HCM} + (\rho_{PCM} - \rho_{HCM})\gamma
\label{eq:interpolation_massDensity}
\end{equation}

\begin{equation}
L_{eff}(\gamma) = \gamma L_{PCM}
\label{eq:interpolation_latentHeat}
\end{equation}

Note that \Cref{eq:interpolation_latentHeat} assumes the latent heat of the HCM is zero. This modeling approximation is justified, given that the HCM serves only to enhance conduction, and the operating temperatures are assumed well below its melting point. Similarly, the melting temperature is fixed to that of the PCM and is not interpolated.

Additionally, the pseudo-densities are regularized using a standard density filter to ensure mesh-independence and preclude checker-boarding. A subsequent threshold projection is applied to the filtered field to obtain a binary design. The specifics of these filtering and projection methods are detailed in \cite{wu2017infill}.

Finally, the proposed optimization framework extends beyond the material layout of conventional TO to include the selection of constituent materials as well. Consequently, the material choices for both the HCM and PCM are treated as design variables. Their representation is detailed in the next section.

\subsection{Differentiable Material Representation}
\label{sec:method_material}

Building upon the density-based design representation, we now detail the representation of discrete material databases in a continuous and differentiable form, amenable to gradient-based optimization. To this end, we employ variational autoencoders (VAEs). VAEs are generative models that have been applied for compression, semi-supervised learning, and interpolation \cite{kingma2019VAE}. In design optimization, they have been used to generate parametric microstructures \cite{padhy2024tomas}, photonic devices \cite{padhy2024photos}, and robust designs \cite{gladstone2021robustTOVAE}. Following the approach of \cite{chandrasekhar2022integrating}, we leverage VAEs' ability to map a discrete database to a continuous, differentiable latent space that can be integrated into a gradient-based optimization framework.

\begin{figure}[H]
	\begin{center}
		\includegraphics[scale=0.3,trim={0 0 0 0},clip]{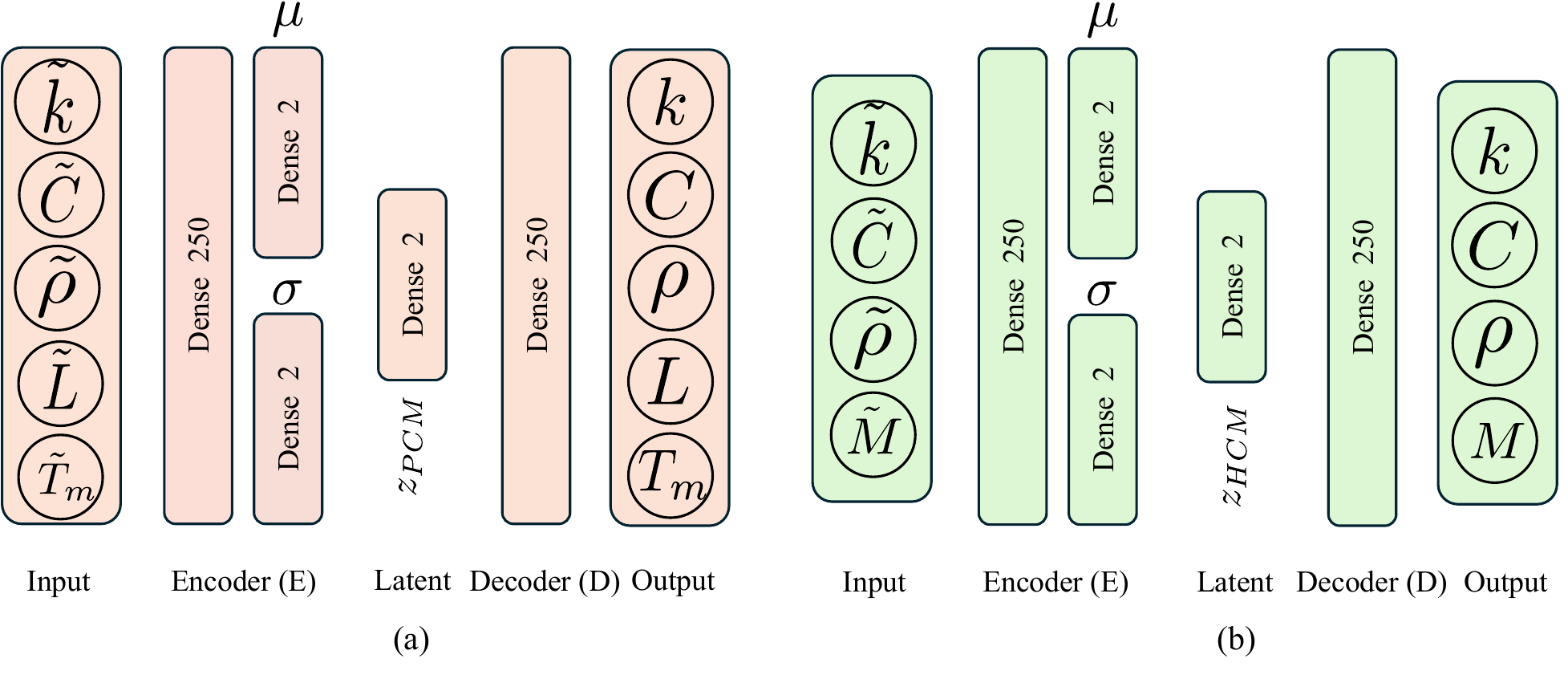}
        \caption{The VAE architecture used to learn a continuous representation for the material properties of (a) the PCMs, and (b) the HCMs.}
        \label{fig:vae_architecture}
	\end{center}
\end{figure}

To this end, we train two distinct VAEs: one for the HCM database and another for the PCM database. As illustrated in \Cref{fig:vae_architecture}, the architecture of each VAE is composed of the following components: (i) an input layer with dimensionality corresponding to the material attributes (five for PCM, four for HCM); (ii) an encoder, $E$, comprising a fully connected network of 250 neurons with ReLU activation functions; (iii) a two-dimensional latent space ($n_L = 2$) parameterized by variables $z_1$ and $z_2$, which is sampled from distributions defined by mean and standard-deviation layers derived from the encoder's output; (iv) a decoder, $D$, also comprising a fully connected network of 250 neurons with ReLU activation functions; and (v) an output layer that reconstructs the original material properties.

The HCM and PCM VAEs are trained on databases containing $n_H = 56$ and $n_P = 140$ materials, respectively. The training process minimizes a loss function composed of a mean-squared-error reconstruction term and a Kullback–Leibler (KL) divergence regularizer. This regularizer encourages the posterior distribution of the latent space to approximate a standard normal distribution, $\mathcal{N}(0,\mathbf{I})$ \cite{kingma2013auto}. The total loss function is given by \Cref{eq:vae_training_loss}, where the terms $\bm{\hat{\Psi}}$ and $\bm{\Psi}$ collectively represent the input and reconstructed material properties, respectively, and the weighting factor $\beta$ is set to $10^{-7}$.

\begin{equation}
    L_v = ||\bm{\Psi} - \bm{\hat{\Psi}} ||_2^2 + \beta\text{KL}(z || \mathcal{N})
    \label{eq:vae_training_loss}
\end{equation}

\begin{figure}[h]
	\begin{center}
		\includegraphics[scale=0.55,trim={0 0 0 0},clip]{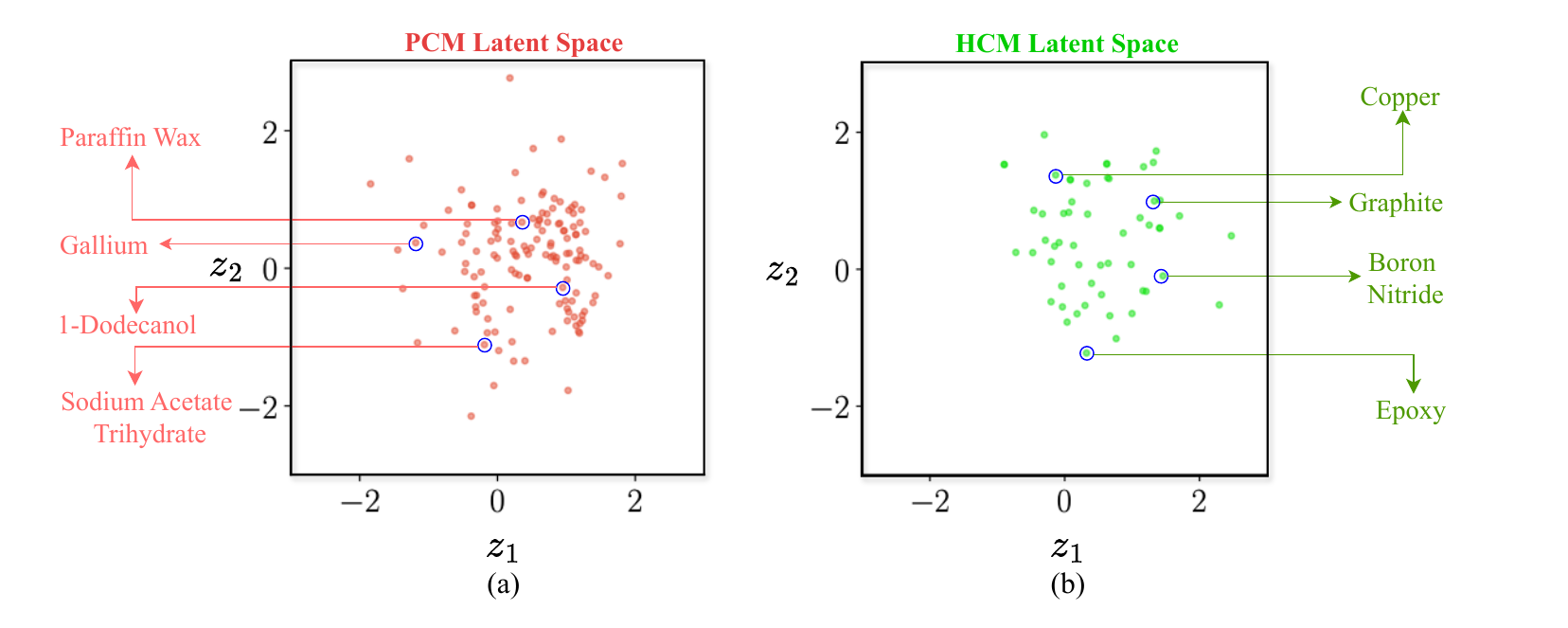}
        \caption{The latent space distribution generated by the (a) PCM-VAE and (b) HCM-VAE.}
        \label{fig:latent_space}
	\end{center}
\end{figure}

\begin{figure}[h]
	\begin{center}
		\includegraphics[scale=0.4,trim={0 0 0 0},clip]{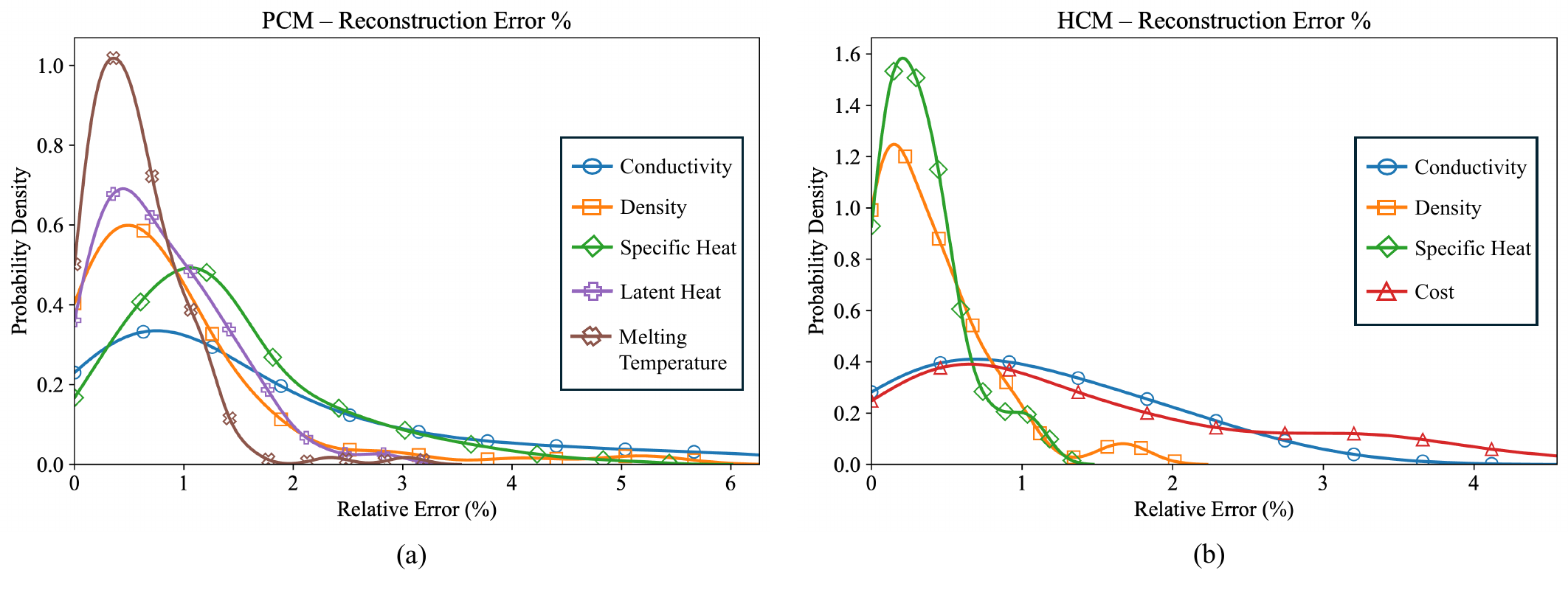}
        \caption{VAE reconstruction error for the properties of (a) PCMs and (b) HCMs.}
        \label{fig:vae_error}
	\end{center}
\end{figure}

The networks were trained for 50,000 epochs using the ADAM optimizer with a learning rate of $2\times10^{-3}$. \Cref{fig:latent_space} visualizes the resulting two-dimensional latent embeddings for both the PCM and HCM databases, with selected materials annotated at their corresponding latent coordinates $(z_1, z_2)$. Furthermore, the mean reconstruction error between the decoded and true material properties is approximately below $8\%$ for both datasets (\Cref{fig:vae_error}) .

Crucially, we note that the VAE-based material representation is inherently differentiable. The decoder network, $D$, parameterizes each material property as a continuous and differentiable function of the latent variables, $\vec{z}$. Consequently, the analytical sensitivities of any material property with respect to $\vec{z}$ can be computed via backpropagation through the decoder. This differentiability enables the integration of the material decoders into the gradient-based design optimization framework, as detailed in the subsequent sections.

\subsection{Finite Element Analysis}
\label{sec:method_fea}

With the HCM and the PCM  materials decoded and the effective thermophysical properties computed for each element based on the supposed material distribution (\Cref{sec:method_designRepMatInterp}), the system's thermal response is evaluated by solving the governing transient, nonlinear heat transfer PDE, given in \Cref{eq:residual_form}. In particular, we employ finite element analysis (FEA) to obtain the temperature distribution and history.

For temporal discretization, we employ an implicit first-order backward-Euler scheme. This yields a discretized system of equations for computing the temperature field at time step $n+1$ as in \Cref{eq:gov_pde_discretized}.

\begin{equation}
  \underbrace{\left( \mathbf{C}_{n+1} + \Delta t \mathbf{K} \right)}_{\tilde{\bm{K}}_{n+1}} \mathbf{T}_{n+1} = \underbrace{\mathbf{C}_{n+1} \mathbf{T}_n + \Delta t\mathbf{F}_{n+1}}_{\tilde{\bm{F}}_{n+1}}
  \label{eq:gov_pde_discretized}
\end{equation}

where, $\bm{T}_n$ is the temperature at the $n^{th}$ time step, $\bm{F}_{n+1}$ is the heat applied at the $n+1$ time step, and $\Delta t$ is the time step size. Furthermore, $\bm{K}$ is the assembled global stiffness matrix, where the elemental stiffness matrix is defined as:

\begin{equation}
  \mathbf{K}^{(e)}(\gamma_e) = \int_{\Omega} (\nabla \mathbf{N})^T \bm{k}_{eff}(\gamma_e) (\nabla \mathbf{N}) \, d\Omega_e
  \label{eq:stiffness_matrix_fea}
\end{equation}

with $\bm{N}$ being the shape function of a four-noded bilinear quadrilateral element in a structured mesh, and $\bm{k}_{eff}$ being the effective thermal conductivity of the element as obtained via \Cref{eq:interpolation_conductivity}. Furthermore,  $\mathbf{C}_{n+1}$ is the assembled global mass matrix, where the elemental mass matrix is defined as:

\begin{equation}
  \mathbf{C}^{(e)}_{n+1}(\gamma_e) = \int_{\Omega} \rho_{eff} \breve{c}_{eff}(\gamma_e, \bm{T}^{(e)}_{n+1}) \mathbf{N}^T \mathbf{N} \, d\Omega_e
  \label{eq:mass_matrix_fea}
\end{equation}

The effective mass density $\rho_{eff}$ and apparent heat capacity $\breve{c}_{eff}$ once again follow the interpolations as discussed in \Cref{sec:method_designRepMatInterp}. \textit {Furthermore, observe that the apparent heat capacity is a function of temperature, making the system of equations nonlinear}. The resulting system of nonlinear residual equations is solved at each time step using a modified Newton-Raphson iterative scheme, as detailed in \cite{alexandersen2023detailed}. To enhance the robustness and accuracy of the numerical solution, a Gradient Galerkin Least Squares (GGLS) stabilization technique is incorporated \cite{pizzolato2018thesis, franca1989galerkin}.

To ensure computational performance, our FEA solver, developed within the JAX ecosystem \cite{jax2018github}, integrates with the high-performance PETSc library \cite{petsc} to solve linear systems at each Newton iteration. Critically, the entire computational framework is constructed to be end-to-end differentiable. This enables the efficient and accurate computation of design sensitivities via automatic differentiation; discussed further in \Cref{sec:method_sensAnalyis}.

\subsection{Optimization}
\label{sec:method_optimization}

Having established our governing physics, design representation, and the material model we now outline the key components of the design optimization framework.

\textbf{Design Variables:} Recall that the optimization jointly determines: (i) the spatial distribution of the HCM and the PCM, and (ii) the selection of the specific HCM and PCM from the material databases. The distribution is determined by designating each element in the discretized domain as HCM or PCM via a vector of pseudo-densities, $\boldsymbol{\gamma} \in \mathbb{R}^{n_e}$.  Concurrently, the choice of the specific HCM and PCM is parameterized by their encoded latent space coordinates, $\mathbf{z}_H \in \mathbb{R}^{n_L}$ and $\mathbf{z}_P \in \mathbb{R}^{n_L}$, respectively. The complete design is thus described by the set of variables $\{\boldsymbol{\gamma}, \mathbf{z}_H, \mathbf{z}_P\}$, for a total of $n_e + 2n_L$ design variables (recall that $n_L = 2$, is the latent-space dimension).

\textbf{Objective:} The primary goal of our framework is to maximize the total energy discharged from the 
LHTES system over a specified time period. We achieve this by minimizing the total stored energy remaining in the system at the end of the discharge cycle. With the effective material properties, and the temperature at an element as $T^{(e)}$, we can express the net elemental energy (sensible and latent) as:

\begin{equation}
h_e(T^{(e)}; \rho_{eff}^{(e)}, c_{eff}^{(e)}, L_{eff}^{(e)}) = \bigg(c_{eff}^{(e)} (T^{(e)} - T_{ref}) + L^{(e)}_{eff} \cdot \Psi(T^{(e)}; T_m + 0.5 \Delta T, \alpha, \Delta T) \bigg) \rho_{eff}^{(e)} v^{(e)}
\label{eq:objective_enthalpy}
\end{equation}

where $T_{ref}$ is a reference temperature, and $v^{(e)}$ is the volume of an element $e$. Then, the objective can be expressed as:

\begin{equation}
    J = \sum\limits_e h_e
    \label{eq:objective}
\end{equation}

\textbf{Cost Constraint:} To ensure a cost-effective design, we impose a cost constraint on the HCM since HCMs are typically more expensive than PCMs. With $\rho(\vec{z}_H)$ being mass density and $M(\vec{z}_H)$ being unit cost of the selected HCM, $v^{(e)}$ being the volume of an element $e$, and $M^*$ being the predefined limit on the total cost, the constraint is given by:
\begin{equation}
    g_m \equiv \frac{1}{M^*}M(\vec{z}_H) \rho(\vec{z}_H) \sum\limits_{e=1}^{N_e}  (1- \gamma_e) v^{(e)} - 1\leq 0
    \label{eq:cost_constraint}
\end{equation}

\textbf{Latent Space Constraint:} Observe that while the optimizer continuously varies the latent coordinates $\vec{z}_H$ and $\vec{z}_P$ during optimization, the materials encoded from the database occupy discrete points in the latent space. In other words, the optimizer is allowed to explore materials outside the database during optimization. However, at convergence, we expect to select materials exclusively from the library. To enforce this, we impose a constraint on the distances between the latent coordinates of selected material instances and those of the database materials. 
With $\bm{\Delta}_{H}$ and $\bm{\Delta}_{P}$ being the pairwise distances between the latent coordinate of the selected material instance and database materials for the HCM and PCM, respectively:

\begin{equation}
    \Delta_{jH} = ||\vec{z}_{H^*}^{(j)} - \vec{z}_H|| \quad , \; j=1,\ldots,n_H
    \label{eq:dist_latent_hcm}
\end{equation}

\begin{equation}
    \Delta_{jP} = ||\vec{z}_{P^*}^{(j)} - \vec{z}_P|| \quad , \; j=1,\ldots,n_P
    \label{eq:dist_latent_pcm}
\end{equation}

where $\vec{z}_{H^*}$ and $\vec{z}_{P^*}$ refer to latent coordinates of points present in material databases.
We can express the combined latent space constraint for the HCM and PCM as:

\begin{equation}
    g_l \;\equiv\; 
    \max\!\Big( \underset{j}{\min}\,(\Delta_{jH}),\; \underset{j}{\min}\,(\Delta_{jP}) \Big) -\epsilon^*
    \;\le\; 0
    \label{eq:latent_dist_cons}
\end{equation}

Note that to facilitate gradient-based optimization, we use the LogSumExp approximations of the min and max functions \cite{zhang2023dive}. Furthermore, we use a continuation scheme on the allowed maximum distance $\epsilon^*$ according to the strategy detailed in \Cref{sec:expts}.

\textbf{Bound Constraints:} Given that the latent space coordinates $\bm{z} \sim \mathcal{N}(0, 1)$, we constrain $\vec{z}_{i} \in [-3, 3]^{n_L} \; , \; i=\{H, P\}$. Further, the design pseudo-density variables are constrained to $\gamma_e \in [0,1] \; \forall e$ with one denoting the element is composed of PCM and zero denoting the element is composed of HCM. 

\textbf{Optimization:} Collecting the objective (\Cref{eq:objective}), PDE (\Cref{eq:residual_form}), and constraints (\Cref{eq:cost_constraint,eq:latent_dist_cons}) the optimization problem can be expressed as:

\begin{subequations}
\label{eq:optimization_base_Eqn}
\begin{align}
\underset{\{ \bm{\gamma}, \bm{z}_H, \bm{z}_P \}} {\text{minimize}} 
           & \quad J \label{eq:optimization_base_objective} \\
R(T) & = 0 \label{eq:opt_residual} \\
g_l &\le 0 \label{eq:constraint_gL} \\
g_m &\le 0 \label{eq:constraint_gm} \\
0 \le \bm{\gamma} &\le 1 \label{eq:constraint_gamma} \\
-3 \le \bm{z}_P, \bm{z}_H &\le 3 \label{eq:constraint_bounds}
\end{align}
\end{subequations}

The constrained minimization problem \Cref{eq:optimization_base_Eqn} is then transformed into an unconstrained loss function minimization, using the log-barrier scheme \cite{kervadec2022logBarrier}. Specifically, the loss function is defined as

\begin{equation}
\mathcal{L} = \frac{J}{J_{0}} + \psi(g_{m}) + \psi(g_{l})
\label{eq:combined_loss}
\end{equation}

where,

\begin{equation}
\psi_{\tau}(g) = \begin{cases}
-\frac{1}{\tau}\ln(-g), \;  & g \le -\frac{1}{\tau^{2}} \\
\tau g - \frac{1}{\tau}\ln(\frac{1}{\tau^{2}}) + \frac{1}{\tau},  \; & \text{otherwise}
\end{cases}
\label{eq:log_barrier}
\end{equation}

and $J_0$ is the initial objective. The constraint penalty parameter $\tau$ is updated using a continuation scheme, making the enforcement of the constraint stricter as the optimization progresses (see \Cref{sec:expts}). The gradient-based Adam optimizer \cite{kingma2014adam} is used to minimize \Cref{eq:combined_loss}. The details for obtaining the gradients are discussed in the next section.

\subsection{Sensitivity Analysis}
\label{sec:method_sensAnalyis}

A critical step in gradient-based topology optimization is the computation of sensitivities, which are the derivatives of the objective and constraint functions with respect to the design variables \cite{padhy2025toflux}. The sensitivity analysis for this work is particularly complex because it must account for transient, nonlinear thermal physics and a computational pipeline that integrates FEA with a neural network (in particular, the material decoders).

To address this complexity, we construct an end-to-end differentiable pipeline using the automatic differentiation (AD) capabilities of the JAX framework \cite{jax2018github}. This allows us to avoid the laborious and error-prone process of manually deriving complex sensitivity expressions. By implementing the entire forward analysis within JAX—from the VAE-based material mapping to the FEA solver—the framework computes the required derivatives to machine precision using reverse-mode AD.

Furthermore, we emphasize two challenges inherent to our physics simulation. The first arises from the use of iterative schemes, such as the Newton-Raphson method, to solve the nonlinear governing equations. A naive application of AD would unroll derivative computation across every solver iteration, a process that is both computationally expensive and memory-intensive. To overcome this inefficiency, we apply the Implicit Function Theorem (IFT) \cite{blondel2022ImplicitDifferentiation, padhy2025toflux}, which enables the direct computation of derivatives from the final converged solution, thereby bypassing the need to backpropagate through the iterative solution history.

The second challenge stems from the transient nature of the simulation, which creates a significant memory bottleneck for the adjoint sensitivity method. An adjoint analysis requires access to the thermal state from all previous time steps to compute the gradient at the current step, and storing this entire state history is often infeasible for long simulations. We mitigate this issue by employing a checkpointing scheme \cite{wang2009minimal, james2015topology}. This technique stores the system's state at select time steps, and during the reverse-time adjoint solve, states between these checkpoints are recomputed on the fly. This method significantly reduces memory requirements at the cost of a moderate increase in computation time, making the analysis of long-duration transient problems feasible.

\subsection{Algorithm}
\label{sec:method_algorithm}
The design optimization procedure consists of two stages: (a) the offline training of the VAEs using the material databases and (b) the iterative design optimization to determine the optimal materials and their distributions.

\subsubsection{VAE Training for Material Representation}

The initial stage focuses on creating a continuous, differentiable, low-dimensional representation of the HCM and PCM material databases.

\begin{enumerate}

\item \textbf{Data Preparation}: We assume that comprehensive material databases for both the HCM and the PCM are available. To account for the different orders of magnitude of the material attributes, each attribute is first log-normalized, followed by rescaling to the range $[0, 1]$ using the minimum and maximum values of each attribute in the datasets. The corresponding scaling parameters are retained for renormalization to recover the physical property values during the optimization phase.

\item \textbf{Network Training}: Using the normalized data, two VAEs—one for the HCM and one for the PCM—are trained as detailed in \Cref{sec:method_material}. Each VAE is trained independently, guided by the loss function in \Cref{eq:vae_training_loss}, until a sufficiently high representational accuracy is achieved.

\item \textbf{Decoder Retention}: After training is complete, the encoders are discarded, while the decoders for the HCM, $D_H$, and PCM, $D_P$, are retained. These decoders act as differentiable surrogates, mapping the continuous latent-space coordinates ($z_H$ and $z_P$) to normalized material properties. These properties are then renormalized using the retained scaling parameters to retrieve the material properties.
\end{enumerate}

\subsubsection{Co-Geometry-Material Optimization}

With the trained decoders, the concurrent optimization of material selection and material distribution is performed. We assume a prescribed design domain, boundary and initial conditions, a maximum allowable cost for the HCM, a target discharge time, and other relevant simulation and optimization parameters. The procedure is as follows:

\begin{enumerate}

\item \textbf{Domain Discretization}: The design domain is discretized into a finite element mesh. This mesh is used for both the thermal analysis and the design parameterization. A pseudo-density design variable is associated with each element.

\item \textbf{Design Variable Initialization}: Elemental pseudo-densities, along with the latent coordinates for the HCM ($z_H$) and PCM ($z_P$)—which serve as the material-choice design variables—are initialized according to the strategy detailed in \Cref{sec:expts}.

\item \textbf{Material Property Decoding}: The latent coordinates, $z_H$ and $z_P$, are passed through their respective trained decoders, $D_H$ and $D_P$. The resulting normalized properties are then renormalized to obtain the physical material properties for the current design iteration \Cref{sec:method_material}.

\item \textbf{Material Interpolation}: Effective material properties for each element are calculated based on the decoded physical properties and the current material distribution, as detailed in \Cref{sec:method_designRepMatInterp}.

\item \textbf{Finite Element Analysis}: Using the calculated material distribution, the nonlinear, transient heat-transfer problem is solved using the finite element method, as discussed in \Cref{sec:method_fea}, to obtain the spatial and temporal distribution of the temperature field.

\item \textbf{Design Objective}: After the thermal simulation, the objective function—the net stored energy at the final discharge time—is evaluated from the temperature history (\Cref{eq:objective}).

\item \textbf{Design Constraints}: The cost constraint (\Cref{eq:cost_constraint}) for the HCM and the latent-space constraint (\Cref{eq:latent_dist_cons}) are evaluated. Bound constraints on the design variables are handled implicitly by the optimization algorithm.

\item \textbf{Sensitivity Analysis}: The sensitivities of the objective and constraint functions with respect to all design variables (the elemental pseudo-densities and the latent coordinates $z_H$ and $z_P$) are computed using automatic differentiation, as discussed in \Cref{sec:method_sensAnalyis}.

\item \textbf{Design Update}: The objective value, constraint values, and their sensitivities are passed to the ADAM optimizer \cite{kingma2014adam}, which then computes an updated set of design variables.

\end{enumerate}

\begin{figure}[h]
	\begin{center}
		\includegraphics[scale=0.55,trim={0 0 0 0},clip]{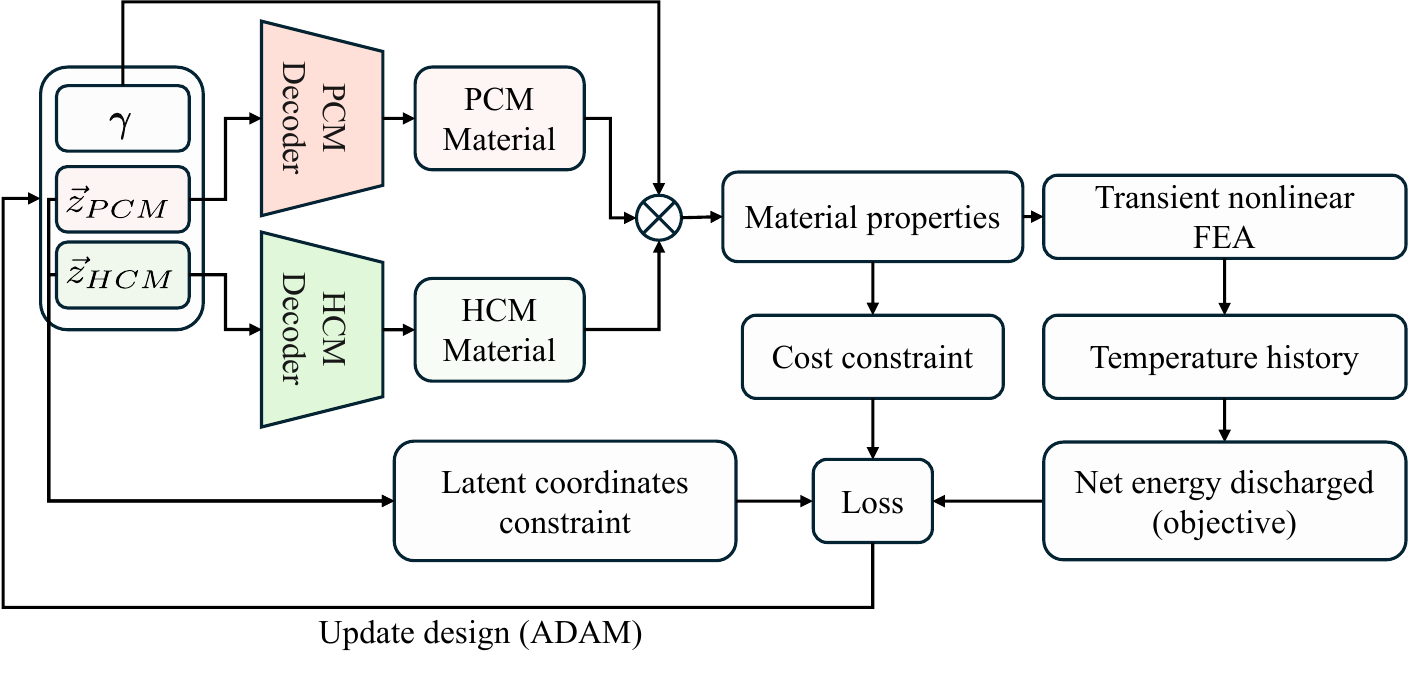}
        \caption{\textcolor{black}{The design optimization procedure.}}
        \label{fig:flowchart}
	\end{center}
\end{figure}

This process (Steps 3–9) is repeated iteratively until the change in design variables between iterations falls below a specified tolerance, or a maximum number of iterations is reached. The optimal material choice and its distribution are obtained as outputs of the optimization. \textcolor{black}{The flowchart of the optimization is shown in \Cref{fig:flowchart}.}

\section{Numerical Experiments}
\label{sec:expts}

In this section, we present several experiments to demonstrate the proposed framework. All experiments were conducted on a MacBook M2 Air, using the JAX library \cite{jax2018github} in Python. Unless otherwise specified, the default parameters for all numerical examples are as follows:

\begin{enumerate}
    \item \textbf{Computational Domain}: The domain (\Cref{fig:domain_bc}(b)) is a quarter-annulus with an inner radius of $0.1 \; m$ and an outer radius of $1.0 \; m$.

    \item \textbf{Discretization}: The domain is discretized using a structured mesh 
    composed of 5000 bilinear quadrilateral elements.

    \item \textbf{Filter}: A convolution type filter with a radius of $0.03 \; m$ is applied to the material pseudo-densities to mitigate checkerboarding and ensure 
    mesh-independence \cite{sigmund1998numericalInstabilites}.

    \item \textbf{Initialization}: The optimization is initialized with a uniform pseudo-density field of $\gamma_e = 0.9$ for all elements. The initial latent coordinates for both the HCM and PCM material decoders are set to $(0,0)$.

    \item \textbf{Thermal Model}: The inner boundary is maintained at a temperature $T_{d}$ of $273 \; K$ (\Cref{eq:residual_form}) while the other boundaries are considered as insulated. The entire domain has a uniform initial temperature of $T_I = 400 \; K$ (\Cref{eq:residual_form}). The reference temperature in \Cref{eq:objective} is taken to be $T_d$. The transient analysis is conducted for 60 time steps with a step size of $800 \; \text{seconds}$ (13.3 hours).

    \item \textbf{Numerical Solvers}: The nonlinear system of equations at each time step is solved using a modified Newton-Raphson method \cite{alexandersen2023detailed} with a convergence tolerance of $10^{-7}$ for the residual norm. The linear system is solved using the PETSc \cite{balay2019petsc} solver.

    \item \textbf{Optimizer}: The optimization is performed using the ADAM optimizer \cite{kingma2014adam} with a 
    learning rate of $5 \times 10^{-2}$.

    \item \textbf{Continuation Schemes}: To promote convergence and smoother optimizations, we adopt continuation schemes (\cite{sigmund1998numericalInstabilites}) for the following parameters:
        \begin{itemize}
            \item The SIMP penalization parameter (\Cref{eq:interpolation_conductivity}), $p$, starts at 1.0 and is increased by 
            0.005 per iteration until it reaches a maximum value of 3.0.
            \item The Heaviside projection parameter starts at 1 and is increased by 0.04 per iteration, up to a value of 64.
            \item The maximum allowed distance in the material latent space $\epsilon^*$ in \Cref{eq:latent_dist_cons}
            starts at 4.0 and is decreased by 0.08 per iteration, until a minimum value of 
            0.02 is reached.
        \end{itemize}

    \item \textbf{Constraints}: The design optimizations are performed with a maximum allowed HCM cost of $\$  600$. The log-barrier parameter $\tau$  in \Cref{eq:log_barrier} is updated at every iteration $'k'$ as $\tau = \tau_0 \mu^k$ with  $\tau_0=3$ and $\mu = 1.02$.

    \item \textbf{Termination Criterion}: The optimization process is run for a maximum of 
    400 iterations.
\end{enumerate}
%---------------------------------%

\subsection{Validation of Hypothesis}
\label{sec:expts_validation}

In this section, we validate the central hypothesis of our work: \textit{the concurrent optimization of material and geometry yields superior LHTES device performance compared to conventional geometry-only optimization or a sequential geometry-first-then-material approach.}

\begin{figure}[h]
 	\begin{center}
		\includegraphics[scale=0.4,trim={0 0 0 0},clip]{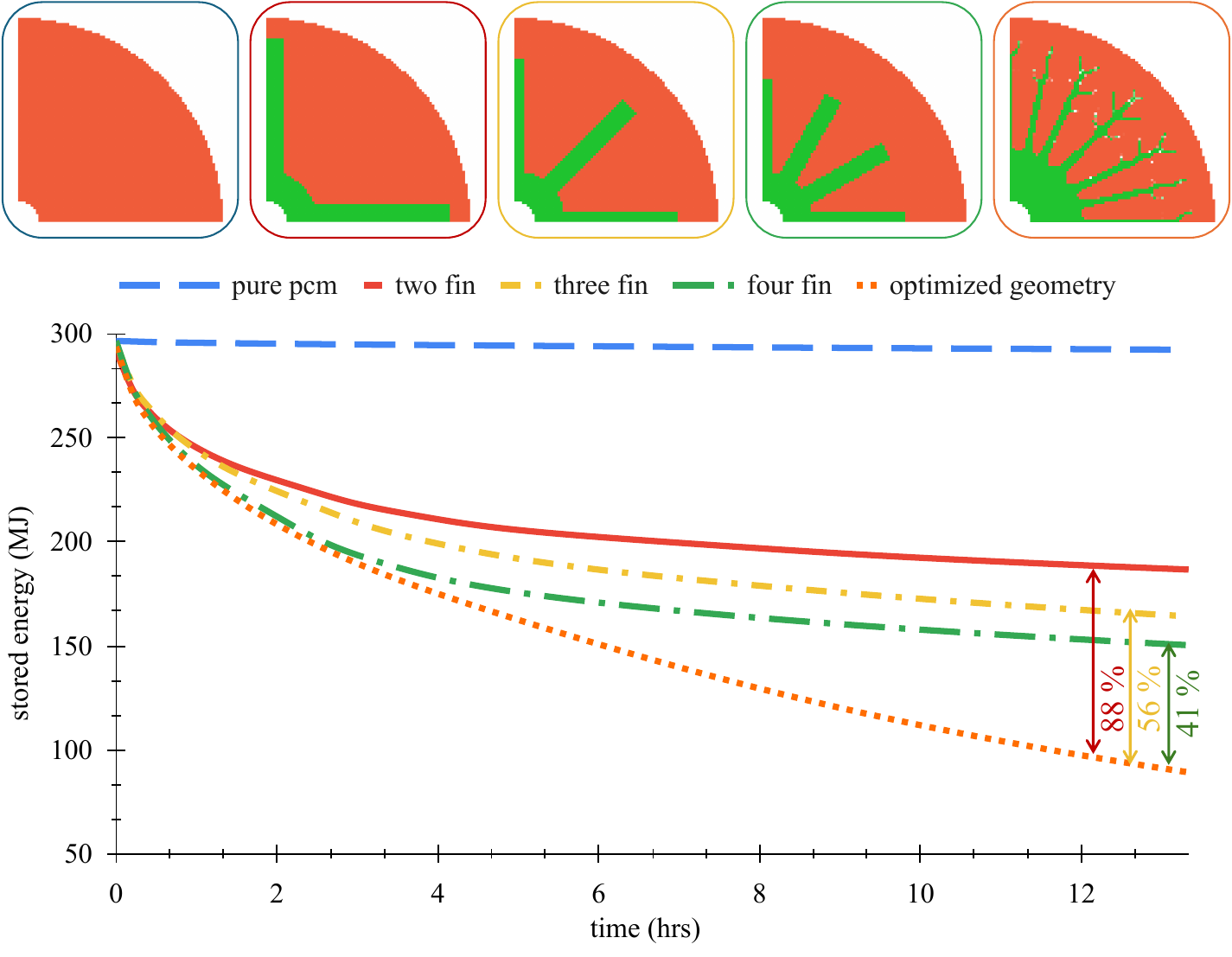}
        \caption{\textcolor{black}{Performance comparison of baseline and geometry optimized designs using fixed materials. The topology-optimized design (dotted orange line) significantly outperforms standard 2-fin, 3-fin, and 4-fin geometries.}}
        \label{fig:verification}
	\end{center}
 \end{figure}

We begin by establishing the advantage of employing TO for the geometric design, with pre-selected materials. We assume commonly used materials: Aluminum as the HCM and RT-25 Paraffin as the PCM. We compare the energy discharge performance of five designs as in \Cref{fig:verification}: a domain filled entirely with PCM, traditional two-fin, three-fin, and four-fin configurations, and the TO design. The pure PCM design serves as a baseline. While it possesses the maximum theoretical latent energy capacity, its poor thermal conductivity limits heat extraction. This results in a low net discharged energy of only $5 \text{ MJ}$ within the discharge period. In contrast, the traditional fin designs, with an allocated $20\%$ of the total volume to the HCM, demonstrate significantly improved performance. Although the inclusion of the HCM reduces the total latent heat capacity, it creates crucial conductive pathways that facilitate more effective heat extraction, thereby increasing the total energy discharged.

To further improve the energy discharged, we now perform TO, maintaining the same $20\%$ HCM volume fraction and material selections. The obtained design outperforms the traditional fin configurations, achieving a $40\%$ to $90\%$ increase in discharged energy. This result underscores the critical role that TO of geometry plays in improving the performance of LHTES systems.

\begin{figure}[H]
 	\begin{center}
		\includegraphics[scale=0.4,trim={0 0 0 0},clip]{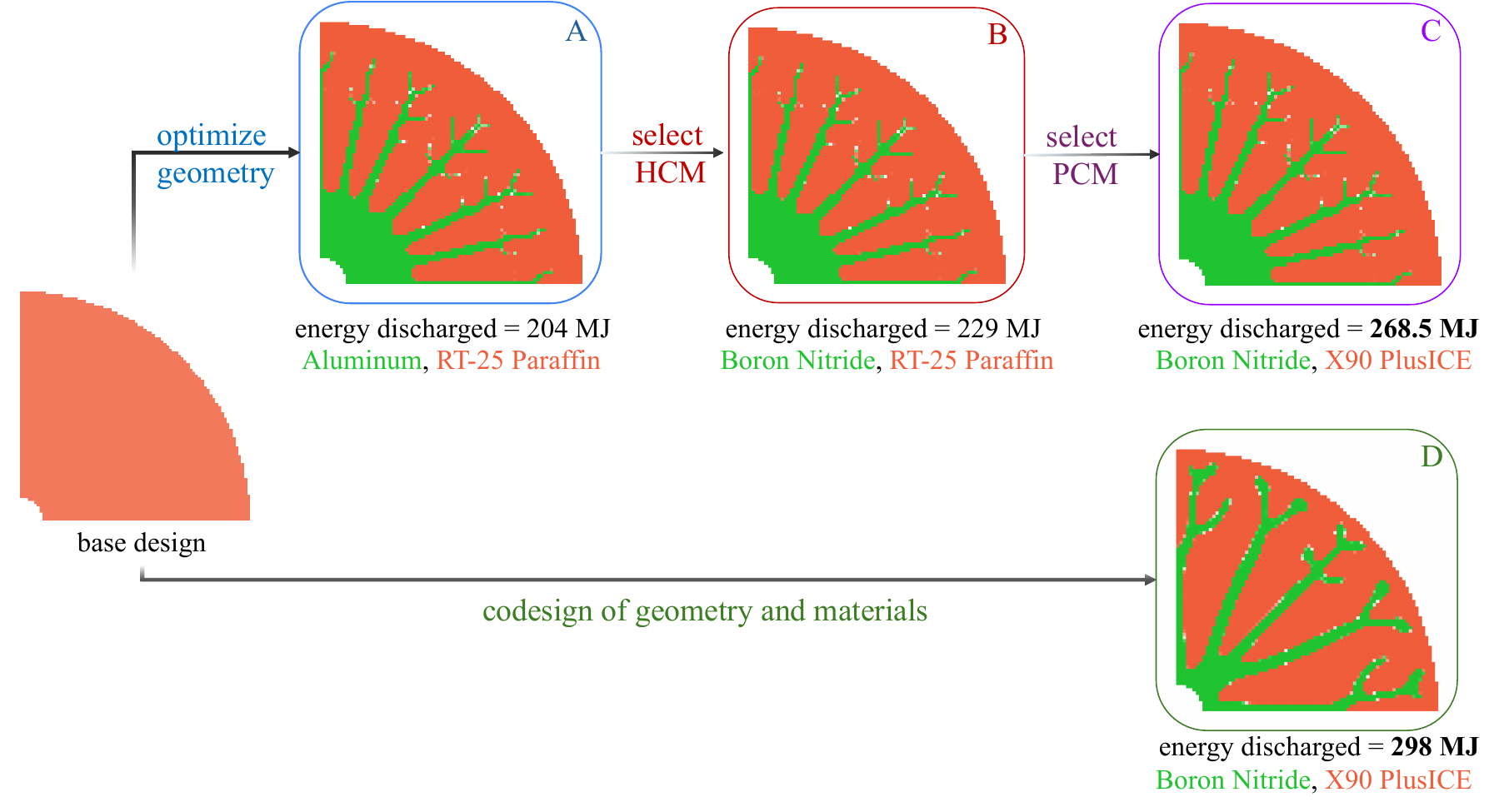}
        \caption{\textcolor{black}{Comparison of the sequential design strategy (top row) versus the proposed concurrent co-design framework (bottom row). The sequential approach optimizes geometry (A), followed by HCM selection (B), and PCM selection (C) as successive steps. In contrast, the co-design formulation (D) concurrently optimizes topology and material choice.}}
        \label{fig:seq_codesign}
	\end{center}
 \end{figure}

\begin{figure}[H]
 	\begin{center}
		\includegraphics[scale=0.4,trim={0 0 0 0},clip]{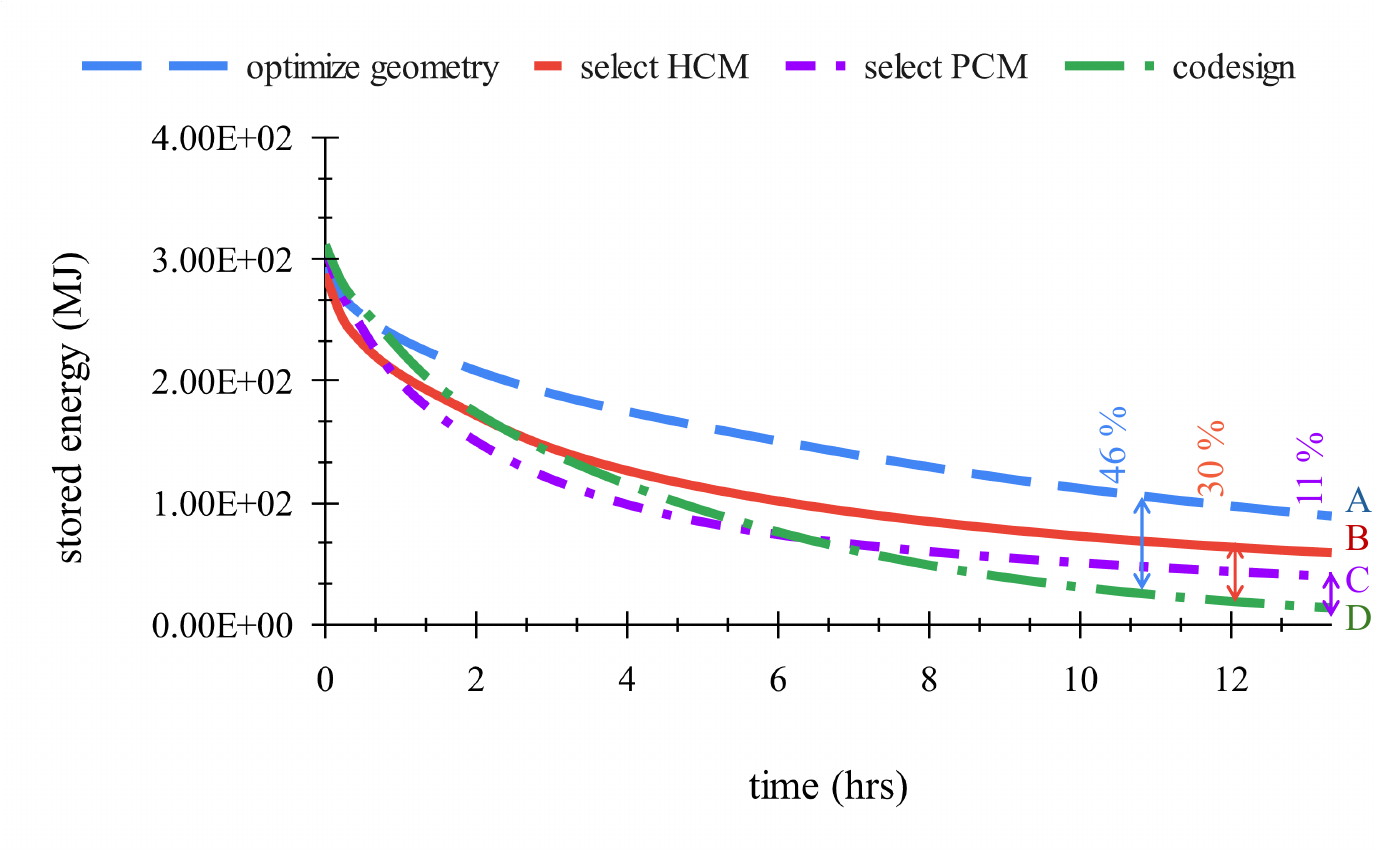}
        \caption{\textcolor{black}{Temporal evolution of the total stored energy (MJ) for different design strategies. The curves illustrate the stored energy remaining in the system over the discharge period for: (A) geometry optimization only, (B) sequential HCM selection, (C) sequential PCM selection, and (D) concurrent co-design.}}
        \label{fig:seq_codesign_plot}
	\end{center}
 \end{figure}

Having demonstrated the importance of geometric optimization, we now compare the sequential design strategy against our proposed concurrent co-design strategy. The results of this comparison are illustrated in \Cref{fig:seq_codesign} and \Cref{fig:seq_codesign_plot}. In the sequential optimization strategy, we use the optimized geometry obtained in the previous step (using pre-selected materials, Aluminum and RT-25 Paraffin). Then, keeping the geometry and PCM fixed, we optimize the HCM choice, which results in the selection of Boron Nitride and an improved energy discharge of $229.5 \text{ MJ}$. Next, using this selected HCM and geometry, we select the optimal PCM. This sequential optimization yields the optimal PCM choice as X90 PlusICE, and a discharge of $268.5 \text{ MJ}$.

In contrast, the concurrent co-design strategy optimizes the geometry, HCM, and PCM simultaneously in a single optimization run. This results in a higher energy discharge of $298 \text{ MJ}$, representing an $11\%$ improvement over the sequential approach. Notably, while the final material combination (Boron Nitride and X90 PlusICE) is the same as in the sequential case, the resulting HCM geometry is markedly different. This highlights the strong coupling between material properties and optimal topology, validating our central hypothesis that concurrent co-design strategy is essential for discovering optimal designs. 

Finally, we note that Boron Nitride, the selected HCM, possesses the highest thermal conductivity in our material database but is also prohibitively expensive (see supplementary material). To address practical design considerations, a cost constraint is introduced in subsequent experiments to balance performance with economic feasibility.

%---------------------------------%
\subsection{Pareto Tradeoff and Convergence}
\label{sec:expts_pareto}

\begin{figure}[H]
 	\begin{center}
		\includegraphics[scale=0.3,trim={0 0 0 0},clip]{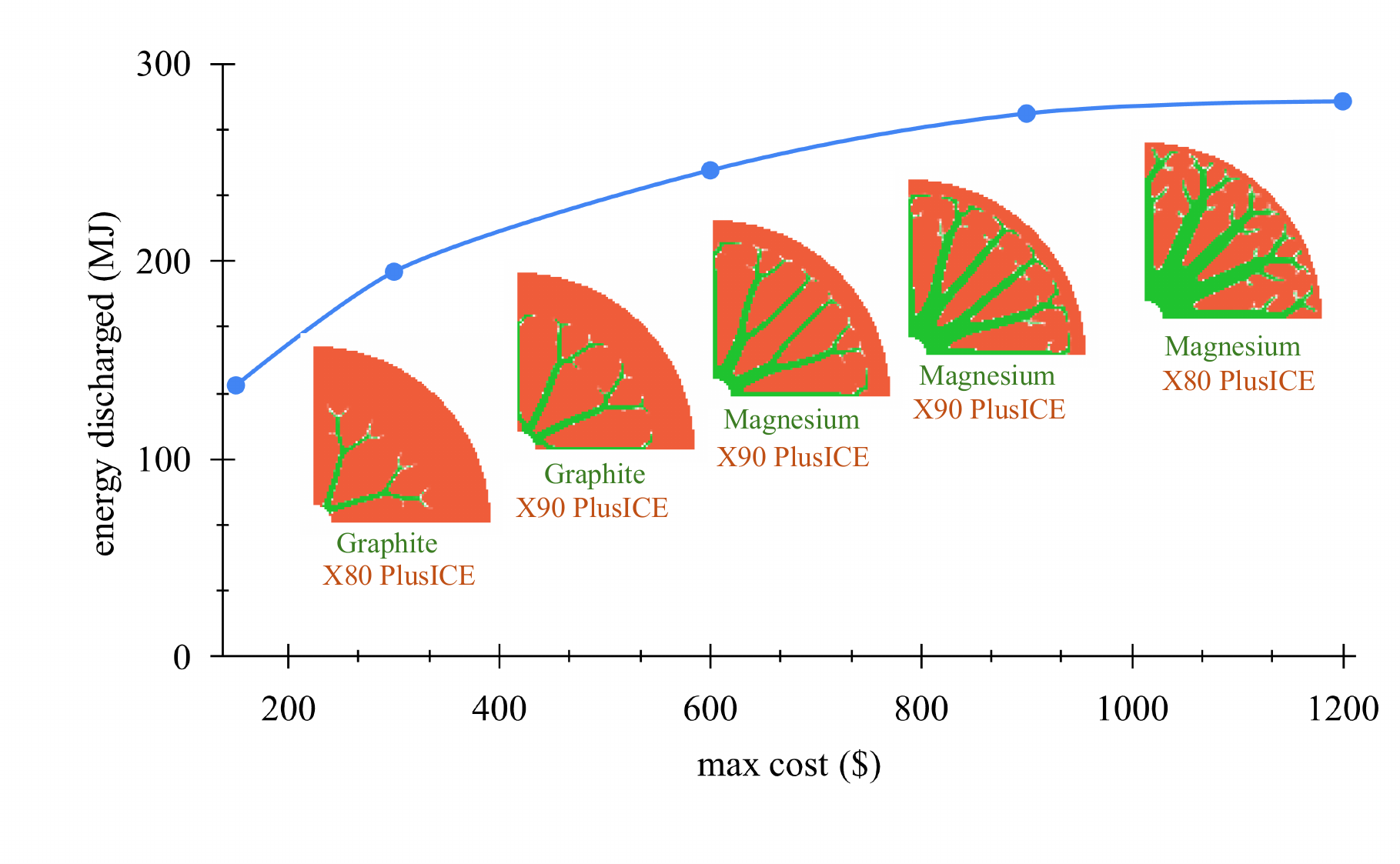}
        \caption{Trade-off between energy discharged and maximum allowable cost.}
        \label{fig:pareto}
	\end{center}
 \end{figure}

A common consideration in engineering design is the trade-off between performance and cost. To demonstrate this, we consider the effect of varying the maximum allowable cost for the HCM. Keeping other parameters including the maximum discharge time, and operating temperatures constant, we perform the co-optimization to obtain the Pareto front as illustrated in \Cref{fig:pareto}. As expected, the discharged energy increases with the allowable cost. 

Interestingly, we note that the optimization selects materials such as graphite and magnesium over more conventional choices such as aluminum. While identifying the exact cause for this selection is challenging—owing to the complex interplay between the HCM and PCM properties, the evolving geometry, and the nonlinear transient nature of the physics; this result highlights an important limitation of our work. For practical implementation, the problem formulation would need to consider additional constraints such as manufacturability, material compatibility, and supply chain availability, alongside the use of more comprehensive and verified material datasets.

\begin{figure}[h]
 	\begin{center}
		\includegraphics[scale=0.4,trim={0 0 0 0},clip]{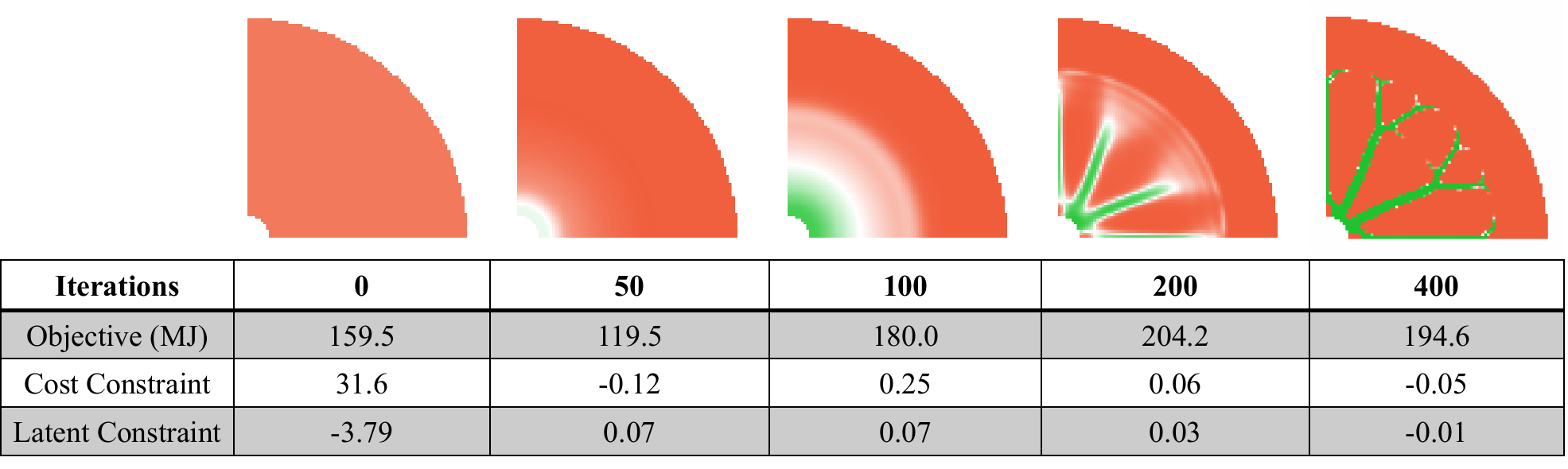}
        \caption{Convergence of the objective and constraints.}
        \label{fig:convergence}
	\end{center}
 \end{figure}

Finally, \Cref{fig:convergence} shows the convergence history for the design corresponding to the $\$ 300$ cost constraint. The objective and constraint values show stable convergence. This optimization was completed in 15 minutes and 32 seconds; similar convergence characteristics were observed for the other experiments.

%---------------------------------%
 \subsection{\textcolor{black}{Effect of Initialization}}
\label{sec:expts_init}
\begin{figure}[h]
 	\begin{center}
		\includegraphics[scale=0.4,trim={0 0 0 0},clip]{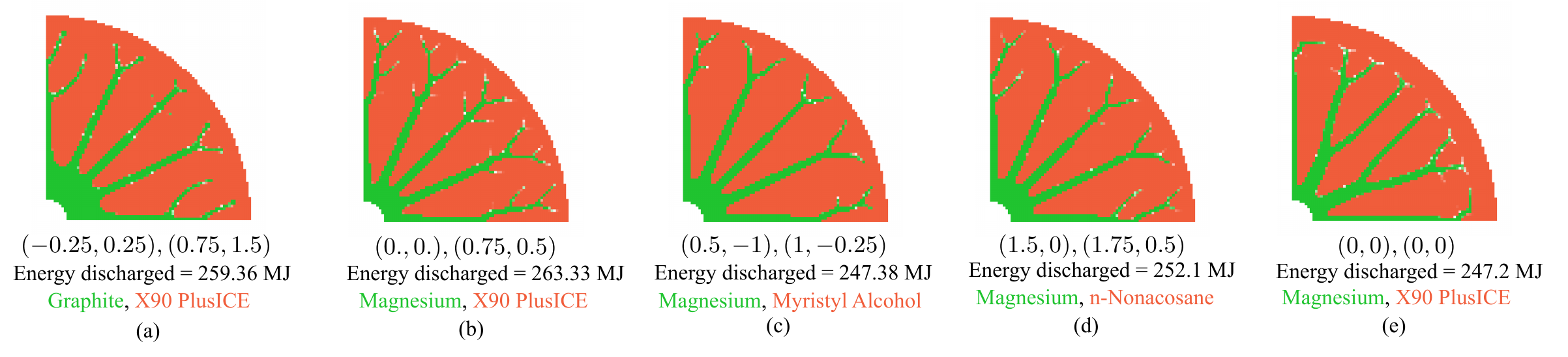}
        \caption{\textcolor{black}{Optimized designs for different initializations. The designs (a-e) illustrate local optima found by the framework. The designs (a-e) show the local optima corresponding to the starting latent positions $\vec{z}_P, \vec{z}_H$ listed below each subfigure for PCM and HCM respectively.}}
        \label{fig:initialization}
	\end{center}
 \end{figure}

\textcolor{black}{
In this experiment, we investigate the influence of the starting design on the optimal design. Given the nonconvex, nonlinear nature of the transient TO problem, we anticipate that different initializations will yield distinct local optima. To study robustness, we performed five independent optimization runs initialized with distinct, randomly sampled points in the material latent space $(\vec{z}_P, \vec{z}_H)$. All optimizations were conducted subject to a maximum cost constraint of $\$600$.\\
\Cref{fig:initialization} illustrates the resulting optimized designs. Observe that while the optimizer converges to diverse geometric topologies and different material combinations, the final performance remains similar (with a maximum deviation of around 4\% from the mean). This suggests that, as expected, the loss landscape contains numerous local solutions with similar performance. Crucially, it demonstrates the robustness of our TO framework, which can discover designs with the desired performance regardless of initialization.\\
Furthermore, a challenge in continuous representation of discrete databases is the physical realizability of the optimized materials, i.e., whether the final solution corresponds to a real material. \Cref{tab:material_verification} compares the material properties predicted by decoders at convergence for design (a) against the values for the nearest materials in the database (Graphite and X90 PlusICE). The decoded properties match the actual physical values with a relative error consistent with \cref{fig:vae_error}. This confirms the effectiveness of the latent space constraint (\Cref{eq:latent_dist_cons}), ensuring the optimizer converges to physically valid materials from the latent space.}

\begin{table}[h]
\centering
\caption{\textcolor{black}{Comparison of decoded material properties at convergence versus actual database values, verifying physical realizability.}}
\label{tab:material_verification}
\begin{tabular}{l|cc|cc}
\hline
\textbf{Property} & \multicolumn{2}{c|}{\textbf{HCM (Graphite)}} & \multicolumn{2}{c}{\textbf{PCM (X90 PlusICE)}} \\
 & \textbf{Decoded} & \textbf{Database} & \textbf{Decoded} & \textbf{Database} \\ \hline
Thermal Conductivity ($W/m\cdot K$) & 200.73 & 200.00 & 0.357 & 0.360 \\
Density ($kg/m^3$) & 2197.06 & 2200.00 & 1190.54 & 1200.00 \\
Specific Heat ($J/kg\cdot K$) & 709.31 & 710.00 & 1539.48 & 1510.00 \\
Latent Heat ($J/kg$) & - & - & 168,390 & 170,000 \\
Melting Temp ($K$) & - & - & 362.97 & 363.00 \\
Cost / Unit (\$) & 1.48 & 1.50 & - & - \\ \hline
\end{tabular}
\end{table}
%---------------------------------%
 \subsection{Variation in Operating Temperature}
 \label{sec:expts_variation_operatingTemp}

The operating temperature range is an important factor for the design of LHTES systems. The PCM's performance depends on its melting point lying within this range. We consider three optimizations with the initial domain temperatures ($T_I$) set to $350 \text{ K}$, $450 \text{ K}$, and $550 \text{ K}$. In each case, the HTF temperature ($T_{d}$) was kept $100 \text{ K}$ below $T_I$.

    \begin{figure}[H]
 	\begin{center}
		\includegraphics[scale=0.5,trim={0 0 0 0},clip]{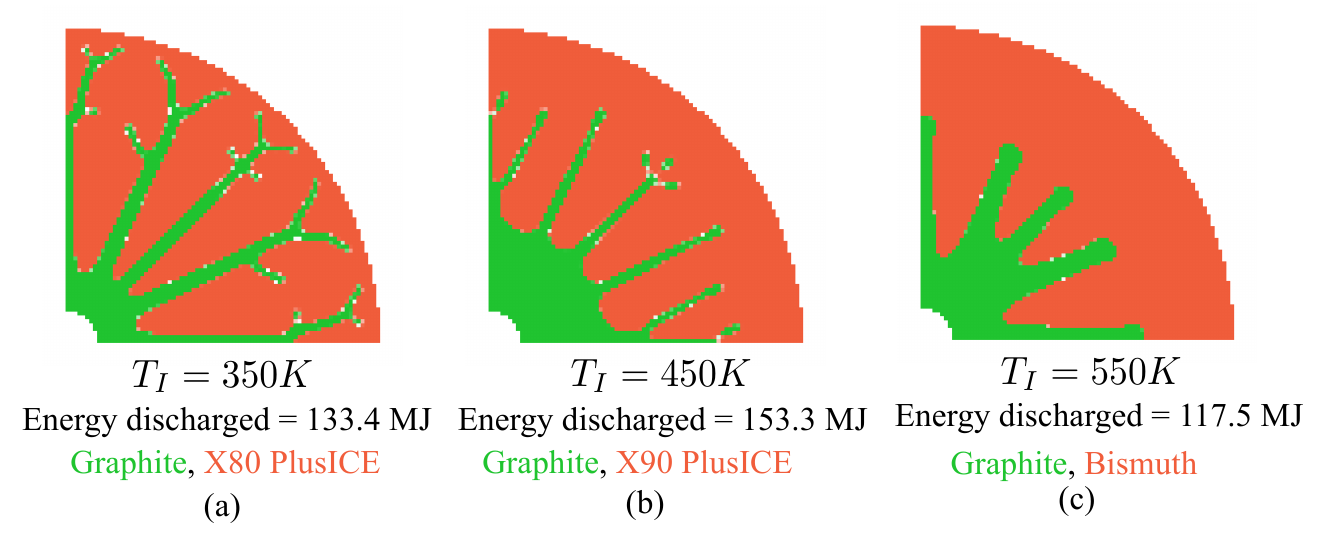}
        \caption{Optimized designs for various initial temperature (a) 350 K, (b) 450 K and (c) 550 K.}
        \label{fig:temp_var}
	\end{center}
 \end{figure}
 
\Cref{fig:temp_var} shows the final optimized designs. The framework selected a different PCM for each operating temperature range. For the $350 \text{ K}$ and $450 \text{ K}$ cases, it chose X80 PlusICE and X90 PlusICE, respectively. For the high-temperature $550 \text{ K}$ case, it chose Bismuth. Each selected PCM has a melting point that falls within its target operating range. This indicates the framework selects appropriate materials. The optimal fin geometry also changes with temperature, once again highlighting the interdependence of geometry and material.

%---------------------------------%

 \subsection{Variation in Discharge Time}
\label{sec:expts_variation_time}

Next, we consider the allowed discharge time for the design of the LHTES systems. We consider four scenarios with different discharge times of 5.5, 11, 16.5 and 22 hours.

\begin{figure}[H]
 	\begin{center}
		\includegraphics[scale=0.5,trim={0 0 0 0},clip]{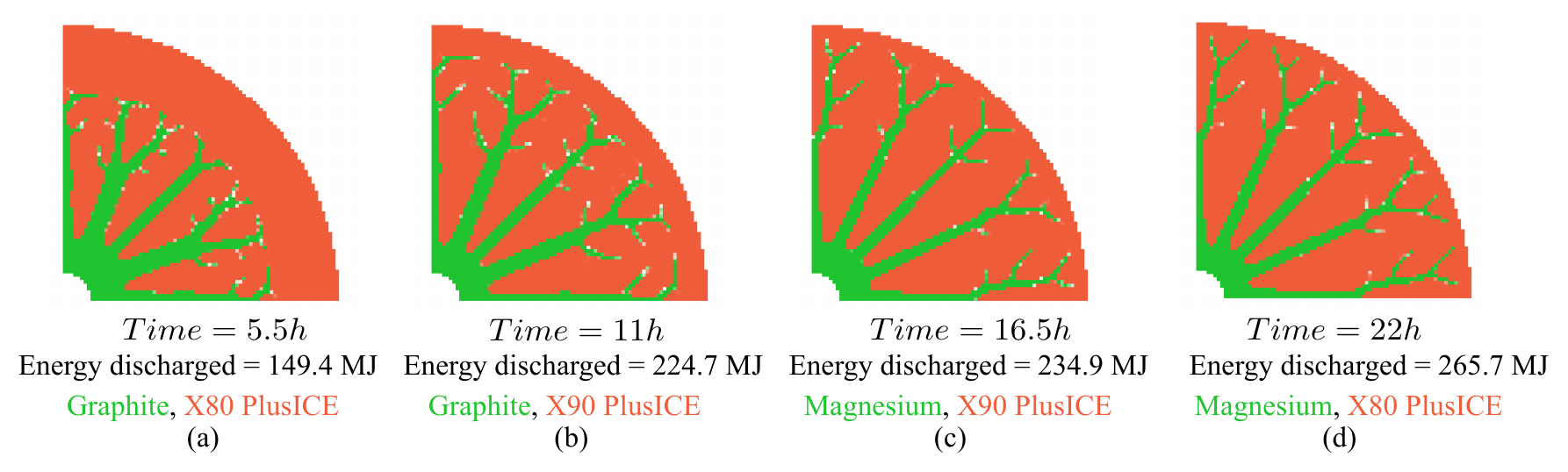}
        \caption{Optimized designs for various discharge times.}
        \label{fig:time_end}
	\end{center}
 \end{figure}

The results are illustrated in \Cref{fig:time_end}. As expected, the total energy discharged increases as the allowed discharge time increases. Furthermore, we observe that the geometry of the HCM fins are thicker and concentrated near the heat source for shorter durations. As the duration extends, the fins branch out more deeply into the PCM domain, creating finer, more complex networks. This allows for more effective heat extraction from regions farther from the source over the longer periods.

Interestingly, the framework also changes the material selection based on the time allowed. For shorter durations ($t = 5.5, 11$ hours), it selects Graphite as the HCM. For longer durations ($t = 16.5, 22$ hours), it switches to Magnesium. This suggests that different materials are optimal for different heat discharge rates and durations.

 %---------------------------------%

\section{Conclusion}
\label{sec:discussion}

This work introduces a computational framework for the co-optimization of geometry and material of latent heat thermal energy storage (LHTES) devices. In particular, we utilize a variational autoencoder (VAE) to represent the discrete material databases of high-conductivity materials (HCMs) and phase-change materials (PCMs) as continuous, differentiable latent spaces (\cite{chandrasekhar2022integrating}). Then, we integrate these differentiable material representations for the simultaneous optimization of material choice and geometric layout. We performed several numerical experiments to validate our central hypothesis: the co-design yields LHTES systems with superior performance compared to conventional sequential design workflows. 

The present study offers several avenues for future research. First, the thermal analysis included only heat conduction, neglecting the effects of convection within the molten PCM. \textcolor{black}{Recent topology optimization studies incorporating natural convection (e.g., \cite{wang2023topology}) have demonstrated that buoyancy forces can influence optimal designs. Specifically, for long charging durations where convection dominates, the optimizer favors simpler V-shaped geometries over complex branching. This geometrical shift occurs to minimize the volume of the HCM (which generally possesses lower specific heat capacity than the PCM), thereby maximizing the total energy storage capacity once convective mixing becomes effective. Extending our differentiable solver to include fluid dynamics is therefore a critical step toward capturing these higher-fidelity trade-offs.} Second, the material databases used in this work were primarily for illustration. The practical implementation of this framework will require the use of comprehensive, experimentally validated material datasets that account for property uncertainties and temperature dependencies. Third, the optimization formulation did not consider constraints such as manufacturability, material compatibility, or cyclic stability, which are essential for real-world deployment. \textcolor{black}{These factors would significantly influence the optimization result. For instance, specific material combinations (e.g., copper and fatty acids) can exhibit chemical degradation or corrosive interaction \cite{ostry2020compatibility}, while galvanic coupling between dissimilar metals (e.g., aluminum and copper) in salt hydrates can lead to structural failure \cite{farrell2006corrosive}. Incorporating compatibility constraints would enable the optimizer to avoid these chemically unstable pairs. Similarly, incorporating constraints on cyclic degradation driven by volumetric expansion \cite{gao2013prediction} and manufacturing limitations \cite{liu2018current} would penalize the complex, thin-walled branching features, thereby favoring robust, printable geometries. Future exploration of this framework will address these practical constraints by incorporating compatibility masks into the latent space and by directly incorporating geometric length-scale constraints into the loss function.} Fourth, we observed that the optimizer can become trapped in local solutions, often requiring multiple initializations. One direction for future work can be to improve global-search robustness and reduce sensitivity to initialization to better approach optimal designs. Fifth, we optimized the LHTES in a 2D domain and focused solely on the selection of the HCM and PCM. As real systems use a circulating HTF, a natural extension can be selecting the optimal HTF and extend the framework to 3D configurations. \textcolor{black}{Sixth, while this study focused on maximizing energy discharge, the underlying differentiable framework is objective-agnostic. Future studies can leverage this flexibility to optimize for alternative objectives, such as minimizing spatial temperature gradients or targeting specific discharge rates, without modifying the core framework.} \textcolor{black}{Finally, future work will target the experimental validation of these designs using Laser Powder Bed Fusion (LPBF) additive manufacturing. As demonstrated recently \cite{morciano2024enhanced}, LPBF is  capable of manufacturing the complex structures predicted by our co-design framework.}

\section*{Appendices}
\appendix
\crefalias{section}{appendix}

\begin{figure}[H]
    \begin{center}
        \includegraphics[scale=0.5,trim={0 0 0 0},clip]{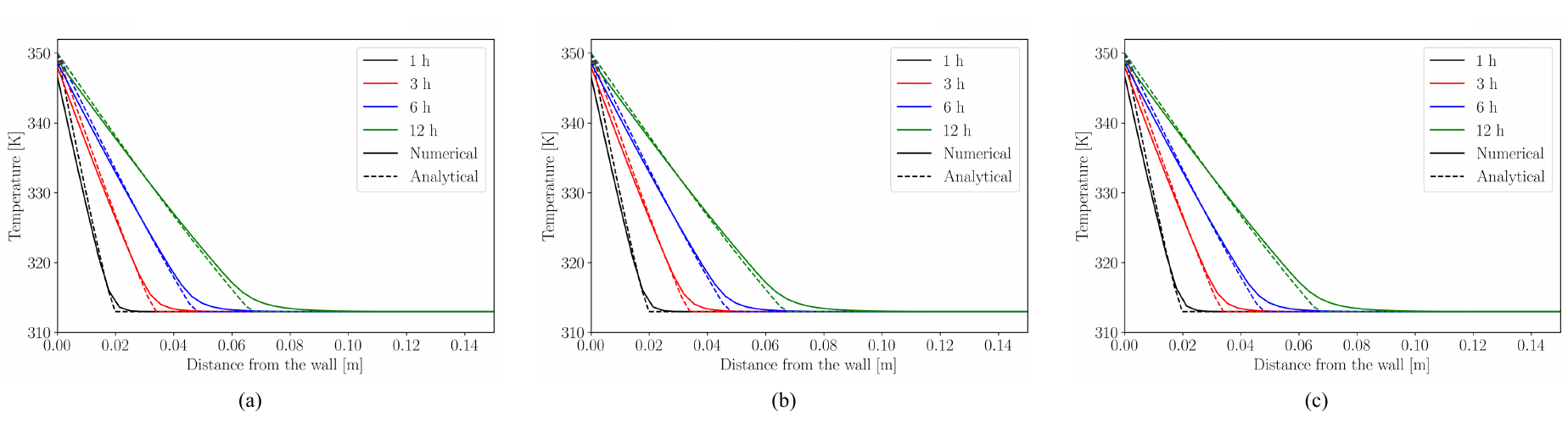}
        \caption{Temporal sensitivity study comparing temperature profiles for time step sizes of $30$ s, $60$ s, and $90$ s with a mesh resolution of $80 \times 80$ elements.}
        \label{fig:stefan_time}
    \end{center}
\end{figure}

\begin{figure}[H]
    \begin{center}
        \includegraphics[scale=0.5,trim={0 0 0 0},clip]{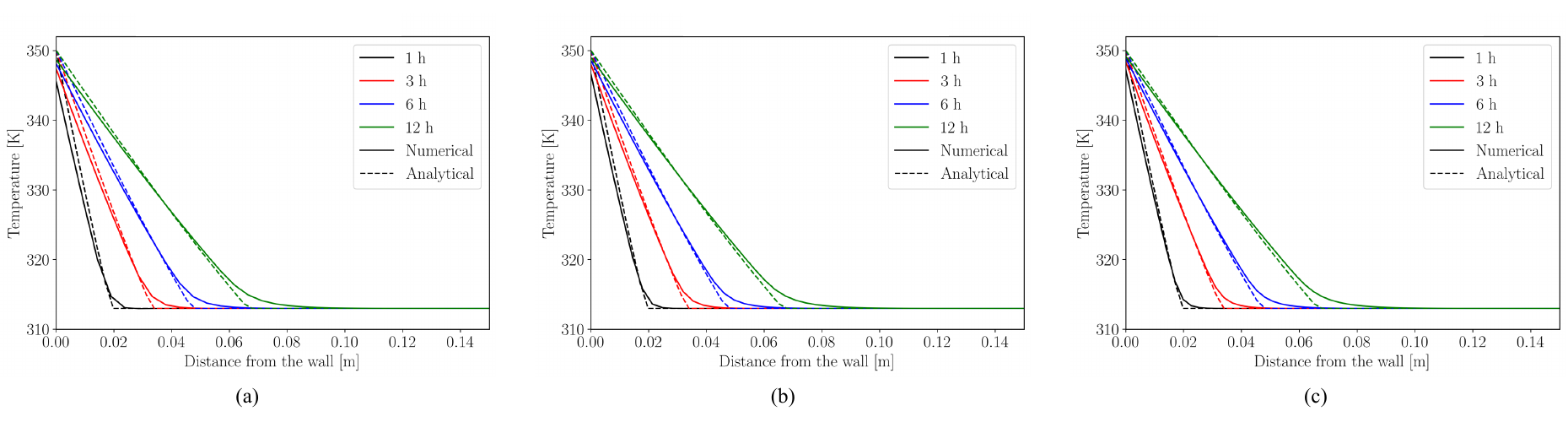}
        \caption{Grid independence study comparing temperature profiles for mesh resolutions of $60 \times 60$, $80 \times 80$, and $100 \times 100$ elements with a fixed time step of $\Delta t = 60$ s.}
        \label{fig:stefan_grid}
    \end{center}
\end{figure}

\textcolor{black}{
\section{Numerical Validation}
\label{sec:appendix_validation}
The numerical accuracy of the computational framework is derived from the finite element solver's implementation of the transient, nonlinear heat equation with phase change. To verify accuracy, we performed a validation study against the classical one-dimensional Stefan problem. This benchmark serves to confirm the solver's ability to capture the moving boundary position and the transient temperature distribution within the PCM.\\\\
The Stefan problem consists of a semi-infinite slab of PCM, initially solid at its melting temperature $T_m$. At time $t=0$, the boundary at $x=0$ is raised to a constant wall temperature $T_w > T_m$, initiating the melting process. Heat transfer is driven solely by conduction in the liquid phase. The problem has an analytical solution for the temperature distribution $T(x,t)$ and the melting front location $\delta(t)$ \cite{ogoh2010stefan}.\\\\
For validation, we utilized the thermophysical properties of Paraffin Wax: density $\rho = 750$ kg/m$^3$, thermal conductivity $k = 0.21$ W/(m$\cdot$K), specific heat capacity $c_p = 2400$ J/(kg$\cdot$K), and latent heat of fusion $L = 175,000$ J/kg. The initial temperature was set to $T_m = 313$ K, and the wall temperature to $T_w = 350$ K. The domain of length $0.28$ m was discretized using a structured mesh of $80 \times 80$ elements, and the time step was set to $\Delta t = 60$ s.\\\\
\Cref{fig:stefan_time} compares the numerical temperature profiles against the analytical solution at various time instances ($t = 1, 3, 6, 12$ hours). The numerical results obtained with a mushy-zone width of $\Delta T = 4$ K show good agreement with the analytical solution. \\\\
To ensure numerical convergence, we conducted  grid independence and temporal sensitivity studies. The grid study evaluated mesh resolutions of $60 \times 60$, $80 \times 80$, and $100 \times 100$ elements while maintaining a fixed time step of $\Delta t = 60$ s (\Cref{fig:stefan_grid}). Conversely, temporal sensitivity was assessed by testing time steps of $30$ s, $60$ s, and $90$ s with a fixed mesh resolution of $80 \times 80$ elements (\Cref{fig:stefan_time}). Similar convergence studies were conducted to determine the appropriate grid resolution and time step size for the main optimization problem.}

\section*{Statements and Declarations}

\section*{Funding and Acknowledgments}
This work was supported by the Department of Mechanical Engineering at the University of Wisconsin-Madison. A.C. conducted this work as a graduate student at the University of Wisconsin-Madison.

\section*{Data Availability and Replication of Results}
The Python code and material data are available at \href{https://github.com/UW-ERSL/TOMATOES}{github.com/UW-ERSL/TOMATOES}

\section*{Author contributions}
R.K.P developed the conceptual framework, wrote the software, conducted the experiments, and prepared the original manuscript. A.C contributed to the development of the conceptual framework, and reviewed the manuscript. K.S supervised the project, acquired funding, and reviewed the final manuscript. A.C. contributed to this work during his tenure as a graduate student at UW-Madison.

\section*{Conflict of Interest}
On behalf of all authors, the corresponding author states that there is no conflict of interest.

\section*{Ethics Approval and Consent to Participate }
Not applicable.

\section*{Declaration of generative AI and AI-assisted technologies in the writing process}

During the preparation of this manuscript, the authors used Google's Gemini and OpenAI's ChatGPT  to improve the language and readability. After using this tool, the authors reviewed and edited the content as needed and take full responsibility for the content of the publication.

\bibliographystyle{unsrt}  
\bibliography{references}

\end{document}